\documentclass[a4paper,11pt]{article}
\pdfoutput=1

\usepackage[english]{babel}
\usepackage{amsmath,amssymb}
\usepackage{jheppub}
\usepackage{amsmath}
\usepackage{amssymb}
\usepackage{color}
\usepackage{graphicx}
\usepackage{hyperref}
\usepackage{xspace,slashed}
\usepackage[utf8]{inputenc}
\usepackage{subcaption}
\usepackage{multirow}
\usepackage{algorithm}
\usepackage{algorithmic}
\usepackage{stackengine}
\usepackage{enumitem}
\usepackage[deletedmarkup=xout]{changes}
\definechangesauthor[name=PN, color=red]{PN}
\definechangesauthor[name=LB, color=blue]{LB}
\definechangesauthor[name=PN, color=pink]{FT}

\usepackage{booktabs}
\usepackage{feynman}

\makeatletter
\g@addto@macro\bfseries{\boldmath}
\makeatother

\newcommand{\as}{\alpha_s}
\newcommand{\mathd}{\mathrm{d}}
\newcommand{\tmop}[1]{\ensuremath{\operatorname{#1}}}
\newcommand\nf{n_{f}}

\newcommand{\xbj}{x_{\rm bj}}
\newcommand{\ellp}{{\ell^\prime}}
\newcommand{\lp}{{l^\prime}}
\newcommand{\Ep}{E^\prime}

\newcommand{\bp}{\beta^\prime}
\newcommand{\tp}{\theta^\prime}
\newcommand{\St}{{\tilde{S}}}
\newcommand{\ax}{{a_x}}
\newcommand{\bx}{{b_x}}
\newcommand{\cx}{{c_x}}
\newcommand{\axi}{{a_\xi}}
\newcommand{\bxi}{{b_\xi}}
\newcommand{\cxi}{{c_\xi}}
\newcommand{\xis}{{\xi^*}}
\newcommand{\ydis}{{y_{\rm dis}}}
\newcommand{\ulp}{{\underline{l}^\prime}}
\newcommand{\barB}{{\bar B}}
\newcommand{\barPhi}{{\bar \Phi}}
\newcommand{\POWHEGBOX}{{\tt POWHEG-BOX}}

\newcommand{\POWHEG}{{\tt POWHEG}}
\newcommand{\Pythia}{{\tt PYTHIA}}
\newcommand{\PythiaEight}{{\tt PYTHIA8}}
\newcommand{\Herwig}{{\tt HERWIG7}}
\newcommand{\SherpaTwo}{{\tt SHERPA2}}
\newcommand{\fastjet}{{\sc FastJet}}

\newcommand{\yad}{{\sc Yadism}}

\definecolor{azure}{rgb}{0.0, 0.5, 1.0}

\newcommand{\NAPaff}{Dipartimento di Fisica Ettore Pancini, Universit\`a di Napoli Federico II and INFN - Sezione di Napoli,
Complesso Universitario di Monte Sant'Angelo Ed.\,6, Via Cintia, 80126 Napoli, Italy}
\newcommand{\TORaff}{Dipartimento di Fisica, Universit\`a di Torino and INFN - Sezione di Torino,
Via Pietro Giuria 1, I-10125 Torino, Italy}
\newcommand{\MUNaff}{Max-Planck-Institut für Physik, Föhringer Ring 6, 80805
München, Germany}
\newcommand{\MILaff}{INFN, Sezione di Milano-Bicocca, and Universit\`a di Milano-Bicocca,
   Piazza della Scienza 3, 20126 Milano, Italy}
\newcommand{\CERNaff}{CERN, Theoretical Physics Department, CH-1211 Geneva 23, Switzerland}

\title{An event generator for
  Lepton-Hadron Deep Inelastic Scattering at NLO+PS with \POWHEG{} including mass effects}

\author[a]{Luca Buonocore,}
\author[b]{Giovanni Limatola,}
\author[c,d]{Paolo Nason}
\author[e,a]{and Francesco Tramontano}
\affiliation[a]{\CERNaff}
\affiliation[b]{\TORaff}
\affiliation[c]{\MILaff}
\affiliation[d]{\MUNaff}
\affiliation[e]{\NAPaff}

\emailAdd{luca.buonocore@cern.ch}
\emailAdd{giovanni.limatola@unito.it}
\emailAdd{paolo.nason@mib.infn.it}
\emailAdd{francesco.tramontano@unina.it}

\preprint{
\begin{flushright}
CERN-TH-2024-070
\end{flushright}
}

\abstract{ We present a generator for lepton nucleon collisions in the
  DIS regime, focusing in particular on processes with a massive
  lepton and/or a massive quark in the final state.  We have built a
  full code matching NLO QCD corrections to parton shower Monte Carlo
  programs in the \POWHEGBOX{} framework. Our code can be used to
  compute NLO+PS accurate fully differential predictions for
  neutral current and charged current
  processes, including processes with an incoming tau
    neutrino, and/or including charm quarks in the final state.
  We also made comparisons with available data and predictions for the
  new neutrino experiments at CERN.  }


\begin{document}

\maketitle


\section{Introduction}\label{sec:intro}
Lepton-hadron Deep Inelastic Scattering (DIS) has been a highly
relevant framework for physics discoveries both for strong and weak
interactions, most noticeably with the discovery of scaling at
SLAC~\cite{PhysRevD.5.528}, and with the discovery of weak neutral currents at
CERN~\cite{GargamelleNeutrino:1973jyy}. Furthermore, DIS is the framework of
choice for the measurement of parton density functions in the proton. Recently,
two new experiments have begun taking data at the LHC, namely the
FASER~\cite{FASER:2023zcr} and SND@LHC~\cite{SNDLHC:2023pun}, that exploit the
large rate of forward neutrinos arising in $p p$ collisions, and promise access
to tau neutrino interactions.\footnote{Up to now, tau neutrinos have been
  revealed at the OPERA~\cite{OPERA:2018nar} and DONUT~\cite{DONuT:2007bsg}
  experiments, which recorded about ten events each.} The study of these
neutrino interactions may have also applications regarding the air
showers~\cite{Reno:2023sdm}.

The upcoming SHiP experiment~\cite{SHiP:2015vad}, a beam-dump experiment
designed for the search of feebly interacting particles, will also study
neutrino interactions. These, together with the plans for the Electron-Ion
Collider (EIC) at BNL~\cite{AbdulKhalek:2021gbh}, and the consideration of
future electron-hadron colliders in Europe, has generated a new interest in DIS
processes also in the theory community.

The DIS cross section for unpolarised Charged Current (CC) neutrino or anti-neutrino scattering
producing an outgoing charged lepton $\ellp$ with mass $m_{\ellp}$,
\begin{equation}
  \label{eqn:DISgeneric}
  \nu/\bar{\nu}(l)+~N(P) \to \ellp/\bar{\ellp}(\lp)+~X(P_X)\,,
\end{equation}
is given by
\begin{equation}
  \label{eqn:HadronicxSec}
  \begin{split}
    \frac{\mathd^2\sigma^{CC}_{\nu /{\bar \nu}}}{\mathd \xbj \mathd y}&=\frac{G^2_FME_\nu}{\pi(1+Q^2/M^2_W)}\biggl\{
    \biggl(y^2\xbj+\frac{m^2_\ellp y}{2E_\nu M}\biggr)F^{CC}_1(\xbj,Q^2)\\
    &+\biggl[\biggl(1-\frac{m^2_\ellp}{4E^2_\nu}\biggr)-\biggl(1+\frac{M \xbj}{2E_\nu}\biggr)y\biggr]F^{CC}_2(\xbj,Q^2)\\
    &\pm\biggl[\xbj y\biggl(1-\frac{y}{2}-\frac{m^2_\ellp y}{4E_\nu M}\biggr)\biggr]F^{CC}_3(\xbj,Q^2)\\
    &+\frac{m^2_\ellp(m^2_\ellp+Q^2)}{4E^2_\nu M^2\xbj}F^{CC}_4(\xbj,Q^2)-\frac{m^2_\ellp}{E_\nu M}F^{CC}_5(\xbj,Q^2)
    \biggr\},
  \end{split}
\end{equation}
where $E_\nu$ is the energy of the incoming neutrino (or anti-neutrino) in the nucleon rest frame, $M_W$ and $G_F$ are the $W$ boson mass and the Fermi coupling constant, and $\xbj, y$ are the usual DIS parameters, defined as
\begin{equation}
  \label{eqn:LIvariables}
  \begin{split}
    Q^2  &=-q^2=-(l-\lp)^2,\\
    \xbj &=\frac{Q^2}{2P \cdot q},\\
    y    &=\frac{P \cdot q}{P \cdot l},\\
    M_X^2&=(l+P-\lp)^2=P_X^2.
  \end{split}
\end{equation}
The contribution of the $F_4$ and $F_5$ structure functions to the cross section
in eq.~\eqref{eqn:HadronicxSec} is suppressed for small lepton masses. This
makes the tau leptons the only viable mean of accessing these two so far
unmeasured structure functions via charged-current tau-neutrino DIS. Moreover,
in the parton model the connection among $F_4$ and $F_5$ and the other structure
functions is straightforward at lowest order, making their prediction very
simple and solid.\footnote{ Albright and Jarlskog in~\cite{Albright:1974ts} find
  that $F_4=0$ and $2xF_5=F_2$, while violations to these relations induced by
  NLO and kinematic mass corrections have been studied in
  ref.~\cite{Kretzer:2002fr}.} A measurement of these two structure functions
would then provide further knowledge about the structure of the proton, and,
beyond that, a further consistency test of the partonic picture through the
verification of their relations with the other structure functions. Furthermore,
a precise measurements of $F_4$ and $F_5$ might be useful to constraint those
scenarios of Beyond Standard Model (BSM) physics at higher scales possibly
related to the leptons of the third generation, that could alter the
contribution of one or another of the form factors to the cross section in
eq.~\eqref{eqn:HadronicxSec} (see for example ref.~\cite{Liu:2015rqa}).

Another relevant phenomenological domain is the production of massive charmed resonances in charged current DIS.
This process can be used to further constraint the uncertainty on proton strangeness (see for example ref.~\cite{Alekhin:2015byh}) that
has an impact on $W$ boson mass extraction at hadron colliders. In fact, at 7 TeV,  approximately 25\% of the inclusive $W$ boson
production rate is induced by at least one second-generation quark, $s$ or $c$, in the initial state~\cite{ATLAS:2017rzl}.
This fraction increases with the center-of-mass energy.

The associated production of a tau lepton and a charm resonance in tau neutrino DIS might be measurable at SND@LHC and SHiP.
The open production of a charm quark and tau lepton with an invariant mass larger than the typical mass of a bottom resonance
could in principle probe new physics scenarios that cannot be explored at B factories.

The status of higher order calculations has reached a remarkable N$^3$LO
accuracy both for DIS structure
functions~\cite{Moch:2004xu,Vermaseren:2005qc,Moch:2008fj,Davies:2016ruz,Blumlein:2022gpp}
and for jet production in DIS~\cite{Currie:2018fgr,Gehrmann:2018odt}, and at
N$^2$LO level for polarised DIS~\cite{Borsa:2022cap}. For the massive cases, NLO
QCD corrections have been calculated since
long~\cite{Gottschalk:1980rv,Gluck:1996ve,Kretzer:2002fr}. The massive structure
functions have been first evaluated in the asymptotic
limit~\cite{Buza:1996wv,Blumlein:2014fqa}, and then at full NNLO
level~\cite{Berger:2016inr}.

As far as fully exclusive Monte Carlo generators are concerned, while the
current status of the calculations for collider processes is at the cutting-edge
with the standard accuracy given by NLO+PS and also NNLO+PS for some classes of
processes, the situation for DIS is less advanced. General-purpose Monte Carlo
events generators, as \Herwig{}~\cite{Bahr:2008pv,Bellm:2019zci},
\SherpaTwo{}~\cite{Gleisberg:2008ta,Sherpa:2019gpd} and
\PythiaEight{}~\cite{Sjostrand:2014zea}, widely used for hadron-hadron
collisions can be also used to simulate massless DIS processes. For the massless
case, a NLO+PS implementation is available within the \Herwig{}
generator~\cite{Carli:2010cg} and, more recently, \POWHEG{} based generators have
been presented in ref.~\cite{Banfi:2023mhz} and ref.~\cite{Borsa:2024rmh} for
the unpolarised and polarised DIS respectively. Furthermore, an NNLO+PS
implementation in the UNNLOPS framework has appeared~\cite{Hoche:2018gti}, still
for the massless case.

A widely used tool for full simulation of neutrino-nucleon interactions is the
generator GENIE~\cite{Andreopoulos:2009rq}. For the case of DIS, it does not
rely on standard procedures for matching fixed-order corrections with a parton
shower. Instead, Born level events are generated according to the higher-order
but inclusive result (i.e. according to the structure functions) and, then,
subsequently processed with \Pythia{}, as discussed in
ref.~\cite{Garcia:2020jwr}. This procedure ensures that the outgoing lepton has
the correct kinematics assuming that the parton shower adopts a recoil scheme
which affects only the coloured partons.

In the present paper we aim to fill the gap, in particular for massive
final states, and we present a full event generator to describe DIS
events for both Neutral and Charged Current interactions (NC and CC
respectively from now on).  In order to do this we first consider the
relevant QCD radiative corrections, then exploit the \POWHEG{}
method~\cite{Nason:2004rx, Frixione:2007vw,Alioli:2010xd} to match a
Next-to-Leading (NLO) fixed order computation to a Shower Monte Carlo
(SMC) program.

The paper is structured as follows. In section~\ref{description} we
outline the basic aspects of our computation, referring to the
appendix for the derivation of the relevant formulae and other
technical details. We also show validation results for both the fixed
order computations and showered event samples.  In section~\ref{pheno}
we present a selection of phenomenologically relevant results and
perform comparisons with available data.  We draw our conclusions in
section~\ref{conclusion}.

\section{Description of the calculation}
\label{description}

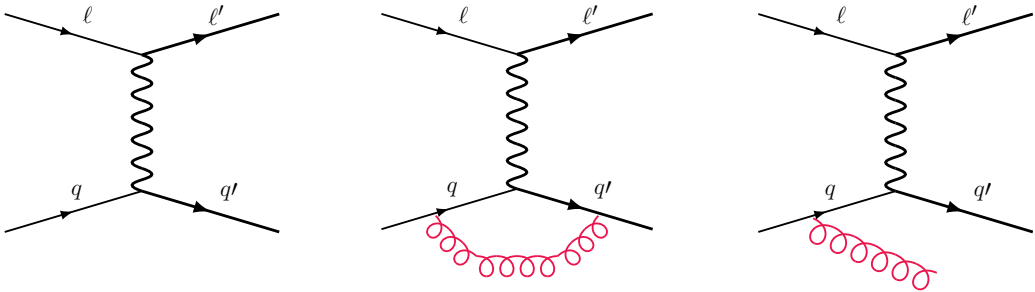
\begin{figure}
  \centering
  \begin{tabular}{p{0.3\textwidth} p{0.3\textwidth} p{0.3\textwidth}}
    \vspace{0cm}
    \resizebox{0.25\textwidth}{!}{%
    \begin{feynman}
      \fermion[lineWidth=3, label={\Huge $q\prime$}]{6.00, 4.60}{8.00, 4.00}
      \fermion[lineWidth=2, label={\Huge $q$}]{4.00, 4.00}{6.00, 4.60}
      \fermion[lineWidth=2, label={\Huge $\ell$}]{4.00, 7.20}{6.00, 6.60}
      \electroweak[lineWidth=3]{6.00, 4.60}{6.00, 6.60}
      \fermion[lineWidth=3, label={\Huge $\ellp$}]{6.00, 6.60}{8.00, 7.20}
    \end{feynman}%
    } &
        \vspace{0cm}
        \resizebox{0.25\textwidth}{!}{%
        \begin{feynman}
          \fermion[lineWidth=3, label={\Huge $q\prime$}]{6.00, 5.00}{8.00, 4.40}
          \fermion[lineWidth=2, label={\Huge $q$}]{4.00, 4.40}{6.00, 5.00}
          \fermion[lineWidth=2, label={\Huge $\ell$}]{4.00, 7.60}{6.00, 7.00}
          \gluon[flip=true, lineWidth=2, endcaps=false, color=eb144c]{4.80, 4.60}{5.40, 4.00}
          \electroweak[lineWidth=3]{6.00, 5.00}{6.00, 7.00}
          \gluon[endcaps=false, flip=true, color=eb144c]{6.60, 4.00}{7.20, 4.60}
          \gluon[flip=true, endcaps=false, color=eb144c]{5.40, 4.00}{6.60, 4.00}
          \fermion[lineWidth=3, label={\Huge $\ellp$}]{6.00, 7.00}{8.00, 7.60}
        \end{feynman}
        }
    &
      \vspace{0cm}
      \resizebox{0.25\textwidth}{!}{%
      \begin{feynman}
        \fermion[lineWidth=3, label={\Huge $q\prime$}]{6.00, 5.20}{8.00, 4.60}
        \fermion[lineWidth=2, label={\Huge $q$}]{4.00, 4.60}{6.00, 5.20}
        \fermion[lineWidth=2, label={\Huge $\ell$}]{4.00, 7.80}{6.00, 7.20}
        \electroweak[lineWidth=3]{6.00, 5.20}{6.00, 7.20}
        \fermion[lineWidth=3, label={\Huge $\ellp$}]{6.00, 7.20}{8.00, 7.80} \gluon[endcaps=false,
        flip=false, color=eb144c]{6.60, 4.00}{4.80, 4.80}
       \end{feynman}%
      }
  \end{tabular}
  \caption{\label{fig:diags} Born, Virtual and Real emission sample diagrams
    for lepton-nucleon DIS, $\ell + N \rightarrow \ellp + X$. }
\end{figure}

In this section we describe the main steps to develop a full event
generator for the simulation of lepton-hadron DIS processes with
NLO+PS accuracy. We start by considering fixed-energy incoming
leptons.\footnote{The case of a broad band beam of incoming leptons,
which is relevant for the application to neutrinos, involves an extra
convolution with the flux of incident leptons as will be discussed in
section~\ref{pheno}.}

The first step corresponds to implement a differential NLO calculation for the
DIS processes. We have re-derived analytic expressions for
all needed Born, Virtual and Real matrix elements for both NC and CC DIS
interactions with massive or massless particles in the final state.
Sample diagrams are shown in figure~\ref{fig:diags}. We have double checked their numerical implementation with
GoSam~\cite{GoSam:2014iqq}.

The second ingredient needed to perform the NLO computation is a subtraction
scheme for the infrared and collinear divergences. The \POWHEGBOX{} framework
implements the Frixione-Kunszt-Signer (FKS)~\cite{Frixione:1995ms} subtraction
scheme that enforces a partition of the real emission phase space according to
the collinear singularities of the real matrix elements. As is the case for any
local subtraction scheme, we need to provide suitable momentum mappings
connecting a Born phase space configuration plus a set of radiation variables to
a real phase space configuration. There is some freedom in the choice of these
mappings. While the particular map has no effect on pure NLO results, it has an
impact when one matches the NLO computation to an SMC program. For the DIS case,
a generic map may introduce some distortions in the distributions of the
leptonic variables that are unnatural when only QCD corrections are considered.

To be more explicit, let us consider for example a CC DIS process with an
initial state lepton scattering off a light quark in the proton producing a
lepton and a massive quark. In the FKS scheme the real emission corrections to
this process has only one initial-state collinear-singular configuration (for
each real subprocess) and no final-state collinear singularities, thanks to the
mass of the final-state quark that acts as a regulator of the collinear
divergence.

We observe that, starting from a Born configuration and a set of radiation
variables, the default initial-state mapping in the \POWHEGBOX{} is built so
that it preserves the invariant mass and the rapidity of the Born final state
partonic system. This choice is particularly suitable for hadron-hadron
collisions with production of massive resonances. The prize is that both initial
state momenta are not preserved. In our case this would imply a real event with
a more energetic incoming lepton with respect to the starting Born
configuration. This leads to an inconsistent formulation of the subtraction
procedure for fixed-energy incoming leptons.\footnote{Note that, considering a
  flux of incoming leptons with variable energies, we can still perform the NLO
  computation following the procedure outlined in section~\ref{pheno} below and
  using the default \POWHEGBOX{} initial-state mapping, even for a very narrow
  energy distribution of the incoming leptons.} Furthermore, at the time of
event generation, the default \POWHEG{} initial-state mapping would also change
the momentum of the final state lepton, and this change will not be modified by
the SMC program, which for the time being is supposed to not alter the momenta
of the leptons. This second problem might alter the value of fiducial cross
sections in passing from the NLO to NLO+PS results when, for example, a cut on
$Q^2$ is applied. When available in the SMC program, one can choose different
recoil schemes to assess the corresponding matching uncertainties, but a
construction that preserves the leptonic momenta seems more justified on a
physical ground.

A first option to overcome, at least in part, the problems mentioned above
consists in building a mapping for initial-state radiation (ISR) that preserves
the energy of the initial lepton and the invariant mass of the Born system. A
second and better solution is provided by a mapping preserving the momenta of
both the initial- and final-state lepton. We remind that general formulae for
such mappings in terms of invariants have been derived since long, see e.g.
ref.~\cite{Catani:1996vz}. Nonetheless, their implementation
within the FKS framework is more involved as one has to adopt the specific
parametrisation of the radiation variables that is employed in the construction
of the FKS counterterms obtained by means of the plus prescriptions.

Similar considerations also apply to final-state radiation (FSR) when one
considers the production of a massless quark at Born level. In this case, the
default mapping implemented in the \POWHEGBOX{} preserves the initial-state
momenta, and so can be directly applied to the case of fixed-energy leptons. On
the other hand, this mapping does not preserve the final-state lepton momentum
as the radiation recoil is globally absorbed by all final-state particles.

For the massless case, ISR and FSR mappings preserving the leptonic momenta have
been derived in ref.~\cite{Banfi:2023mhz}. We have derived their generalisation
for a massive quark and lepton in the final state. We have verified that our
formulae smoothly reduce to the ones in ref.~\cite{Banfi:2023mhz} when
approaching the massless limit. Since the construction of these mappings is
rather technical and the corresponding formulae quite lengthy, we report them in
appendix~\ref{sec:DIS_mapping}.

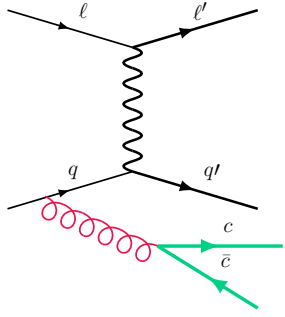
\begin{figure}[t]
  \centering
    \resizebox{0.25\textwidth}{!}{%
    \begin{feynman}
      \fermion[lineWidth=3, label={\Huge $q\prime$}]{6.00, 6.20}{8.00, 5.60}
      \fermion[lineWidth=2,label={\Huge $q$}]{4.00, 5.60}{6.00, 6.20}
      \fermion[lineWidth=4, color=00d084, label={\Huge $c$}]{6.40, 5.00}{8.40, 5.00}
      \fermion[lineWidth=2, label={\Huge $\ell$}]{4.00, 8.80}{6.00, 8.20}
      \fermion[lineWidth=4, color=00d084, label={\Huge $\bar{c}$}]{8.00, 4.00}{6.40, 5.00}
      \electroweak[lineWidth=3]{6.00, 6.20}{6.00, 8.20}
      \fermion[lineWidth=3, label={\Huge $\ellp$}]{6.00, 8.20}{8.00, 8.80}
      \gluon[endcaps=false, flip=false, color=eb144c]{6.40, 5.00}{4.60, 5.80}
    \end{feynman}%
}
\caption{\label{fig:diags_glspl_ccb} Sample diagram of charm-anti charm-pair
  production in gluon splitting.}
\end{figure}

Before concluding this section, we make a comment on charm production in CC DIS.
Beside the EW production mechanism considered in this work, which is sensitive
to the strange content of the proton, charm (anti-)quarks can be produced also
in final-state gluon splitting processes, as depicted in
figure~\ref{fig:diags_glspl_ccb}.
The latter processes can be enhanced by the valence densities and can compete
with the EW production. We observe that starting from NNLO, the distinction
between the two production mechanisms becomes less clear. Nonetheless, working
at NLO and in a scheme in which the charm is treated as a massive quark, as we
do, the gluon splitting process is finite and can be treated separately. For the
rest of the work we focus on the EW production mechanism only.

\subsection{NLO validation}
\label{sec:nloval}
We have compared our \POWHEG{} implementation of the NLO corrections to an
explicit calculation of the inclusive double differential cross section formulae
(see for example eq.~\eqref{eqn:HadronicxSec} for the case of CC neutrino and
anti-neutrino scattering). The relevant proton structure functions are obtained
through the convolution of the parton densities with the NLO coefficient
functions taken from Ref~\cite{Furmanski:1981cw} and ref.~\cite{Kretzer:2002fr}
for the massless and massive case respectively.

\begin{figure}
  \includegraphics[width=0.45\textwidth,page=12]{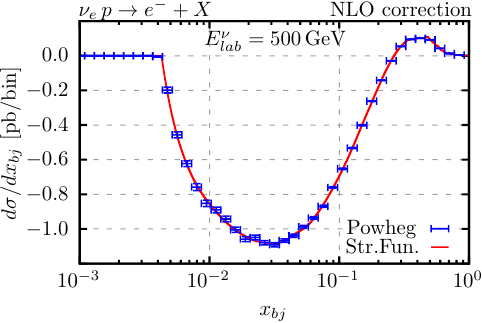}
  \includegraphics[width=0.45\textwidth,page=12]{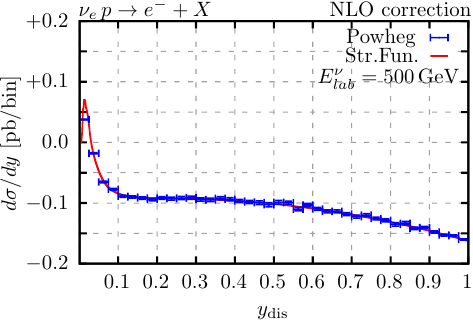}
  \includegraphics[width=0.45\textwidth,page=8]{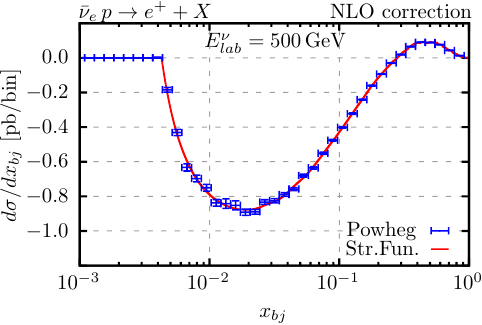}
  \hspace{1.25cm}
  \includegraphics[width=0.45\textwidth,page=8]{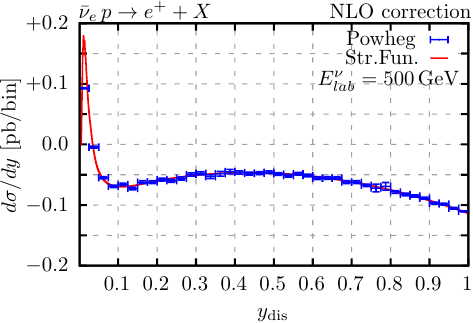}
    \caption{\label{fig:CCnxy}
    Order $\as$ contributions to Björken $\xbj$ (left panels) and inelasticity $\ydis$ (right panels) for
    CC $\nu_e$ (top panels) and $\bar{\nu}_e$ (bottom panels) DIS.
    We considered incoming $\nu_e$ ($\bar{\nu}_e$) with fixed energy equal to 500 GeV in the
    nucleon rest frame.
    We have used the NNPDF31\_nlo\_as\_0118\_nf\_4 PDFs, set  $\mu^2_{\rm{R}}=\mu^2_{\rm{F}}=Q^2$,
    and applied the kinematic cut $Q^2 > 4\,\rm{GeV}^2$.
    The blue points are obtained with our implementation of DIS in the \POWHEGBOX{}
    while the red solid curve is the result of the structure function calculation.
  }
\label{fig:nCC}
\end{figure}

\begin{figure}
  \includegraphics[width=0.45\textwidth,page=11]{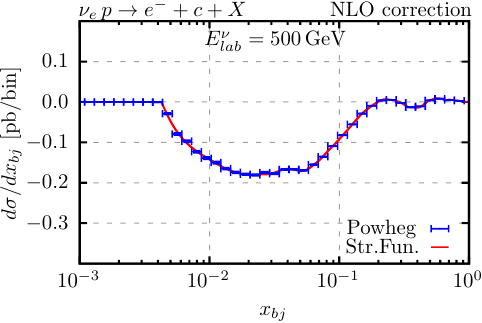}
  \includegraphics[width=0.45\textwidth,page=11]{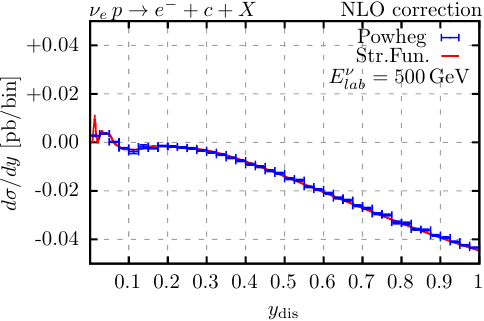}
  \includegraphics[width=0.45\textwidth,page=7]{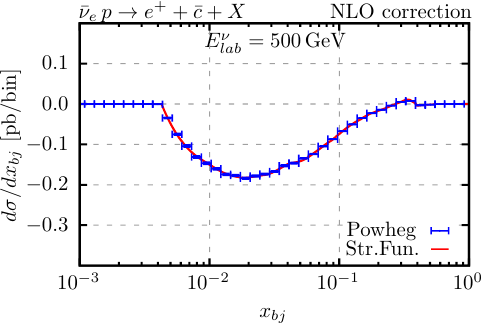}
  \hspace{1.25cm}
  \includegraphics[width=0.45\textwidth,page=7]{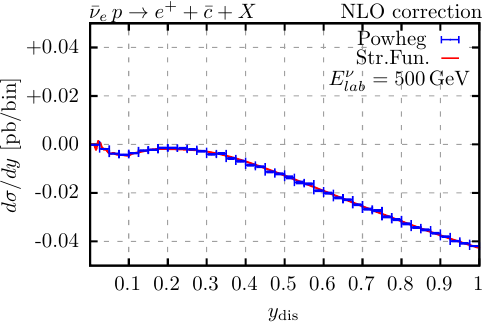}
  \caption{Same as figure~\ref{fig:CCnxy} for charged current $\nu_e$ ($\bar{\nu}_e$) DIS
    with charm quark production setting $m_c = 1.5\,$GeV.
  }
\label{fig:nCCc}
\end{figure}

\begin{figure}
  \includegraphics[width=0.45\textwidth,page=14]{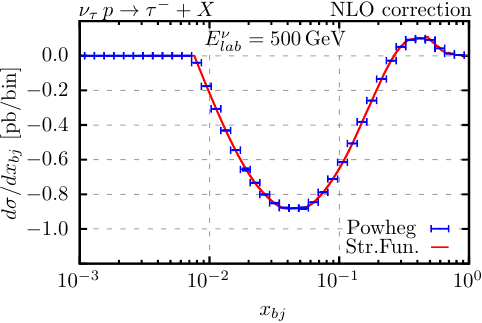}
  \includegraphics[width=0.45\textwidth,page=14]{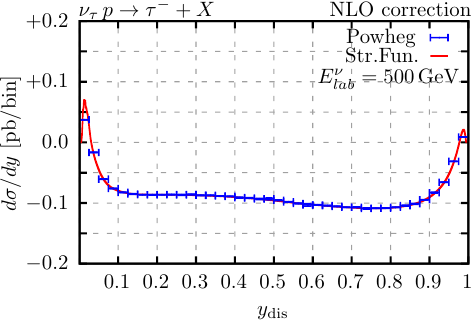}
  \includegraphics[width=0.45\textwidth,page=10]{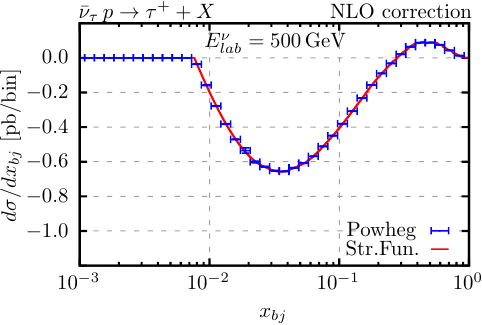}
  \hspace{1.25cm}
  \includegraphics[width=0.45\textwidth,page=10]{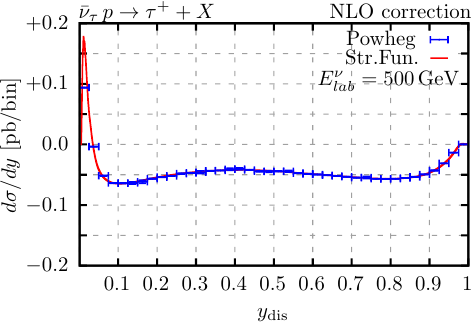}
    \caption{Same as figure~\ref{fig:CCnxy} for charged current $\nu_\tau$ ($\bar{\nu}_\tau$) DIS
      with  $m_\tau = 1.777\,$GeV.
  }
\label{fig:nCCtau}
\end{figure}

Since for many cases and in many kinematic regions the corrections are rather
small, for the purpose of validation, in Figure~\ref{fig:nCC}, \ref{fig:nCCc}
and \ref{fig:nCCtau} we show only the $\alpha_s$ contributions, i.e. for each
distribution we plot the difference NLO\,--\,LO. To generate validation plots we
have considered incoming leptons with fixed energy equal to 500 GeV in the
nucleon rest frame and used the {\tt NNPDF31\_nlo\_as\_0118\_nf\_4}
PDFs~\cite{NNPDF:2017mvq} with $\as =0.118$ and $\nf=4$ through the {\tt LHAPDF}
interface~\cite{Buckley:2014ana}. For the renormalisation ($\mu_{\rm{R}}$) and
factorisation ($\mu_{\rm{F}}$) scales we set
$\mu^2_{\rm{R}}=\mu^2_{\rm{F}}=Q^2$. We also apply the kinematic cut
$Q^2 > 4\,\rm{GeV}^2$ to stay in the DIS regime where perturbation theory is
well applicable.

In particular, in Figure~\ref{fig:nCC} we show the Björken $\xbj$ and
inelasticity $\ydis$ distributions for $\nu_e$ and $\bar{\nu}_e$ CC DIS. The
blue points are obtained with our implementation of DIS in the \POWHEGBOX{}
while the red solid curve is the result of a direct numeric integration of DIS
formulae given in terms of proton structure functions, as in
eq.~\eqref{eqn:HadronicxSec}. This has been obtained using standard gaussian
quadrature routines to perform the convolution of the parton densities with the
coefficient functions.

In Figure~\ref{fig:nCCc} and~\ref{fig:nCCtau}, the same distributions are shown
for the cases of $\nu_e$ (or $\bar{\nu}_e$) CC DIS producing a charm quark ($m_{c}=1.5\,$GeV), and
$\nu_\tau$ (or $\bar{\nu}_{\tau}$) CC DIS producing a tau lepton ($m_{\tau}=1.777\,$GeV),
respectively. For all initial and final state configurations, we found perfect
agreement between the fixed-order results obtained with the \POWHEG{} generator
and the ones obtained with the direct calculation using the structure functions.
We have also checked that the same level of agreement is achieved for all
choices of the available momentum mappings employed in the subtraction
procedure. This represents a non-trivial test of the newly derived DIS momentum
mappings presented in appendix~\ref{sec:DIS_mapping}. Additional sets of
validation plots are reported in appendix~\ref{sec:NLO_val_plot} for
completeness.

\subsection{Event generation and impact of radiative corrections}
\label{sec:validatio-LHE}

The first emission is generated according to the \POWHEG{} master
formula~\cite{Frixione:2007vw}
\begin{equation}\label{eq:radmaster}
  d\sigma_{\textrm{NLO}} =\barB(\Phi_{B}) d\Phi_{B} \left[\Delta_{\textrm{NLO}}(\Phi_{B},t_{\textrm{min}}) + \sum_{\alpha}  \frac{\left[ d \Phi_{\textrm{rad}} \Delta_{\textrm{NLO} }(\Phi_{B},K_{T}(\Phi_{R}))R(\Phi_{R})  \right]^{\barPhi_{B}^{ \alpha}=\Phi_{B}}_{\alpha}}{B(\Phi_{ B})} \right]\;,
\end{equation}
where $\Phi_B$ is the Born phase space and $\Phi_R$ is the real phase space
(that includes the radiation of one extra parton). $\Phi_R$ is mapped
biunivocally into a Born and a radiation phase space, so that
$\mathd \Phi_R=\mathd \Phi_B\,\mathd \Phi_{\rm rad}$. In the above equation, $B$
and $R$ are, respectively, the Born and real squared matrix elements
averaged/summed over colors and spins, $\barB$ entails the NLO corrections
inclusively integrated over the radiation phase space. The sum runs over all the
singular regions, labeled by $\alpha$, and the \POWHEG{} Sudakov reads

\begin{equation}\label{eq:pwgSudakov}
  \Delta_{\textrm{NLO} }(\Phi_{B},p_{T}) = \Theta(p_{T}-t_{\textrm{min}}) \exp\left\{-\sum_{\alpha}\int \frac{\left[ d \Phi_{\textrm{rad}}  R(\Phi_{R}) \right]^{\barPhi_{B}^{ \alpha}=\Phi_{B}}_{\alpha}}{B(\Phi_{B})} \Theta\left( K_{T}(\Phi_{R}) - p_{T} \right) \right\}\;.
\end{equation}
According to eq.~\eqref{eq:radmaster}, resolved radiation is generated with a
hard scale, given by the $K_T$ function, down to some characteristic hadronic
scale $t_{\textrm{min}}$, which in \POWHEG{} is chosen to be
$t_{\textrm{min}}=0.8\;\textrm{GeV}^{2}$.

The evolution variable $K_{T}(\Phi_{R})$ is a smooth function of the radiation
variables, which is required to approach the transverse momentum of the
radiated parton near the soft and collinear limits. For ISR, assuming
that the incoming lepton is moving along the positive $z$-direction, we adopt
the definition
\begin{equation}
  K^{2}_{T}(\Phi_{R}) \equiv  K^{2}_{T}(\Phi_{B},\Phi_{\text{rad}}) = \frac{{\bar s}}{2}\frac{\xi^{2}(1+y)}{1+\xi y}
\end{equation}
where ${\bar s}$ is the CM energy of the underlying Born configuration
$\Phi_{B}$, $\xi$ is twice the ratio of the radiated-parton energy over the
partonic CM energy, and $y$ is the cosine of the angle of the radiated parton
with respect to the positive $z$ axis in the partonic CM frame. Notice that the
collinear limit, in this specific case, is given by $y\to-1$. The term $1+\xi y$
in the denominator, which reduces to $1$ in the soft limit, gives the correct
behavior in the limit of a hard and collinear emission.\footnote{We note that
  our choice of $K_{T}$ in the ISR case differs from the one in
  Ref~\cite{Banfi:2023mhz} by subleading terms. }

For FSR, we adopt the definition
\begin{equation}
  K^{2}_{T}(\Phi_{R}) \equiv  K^{2}_{T}(\Phi_{B},\Phi_{\text{rad}}) = \frac{{\bar s}}{2}\xi^{2}(1-y)
\end{equation}
where, now, $y$ is the cosine of the angle between the radiation and the emitter
parton. The interested reader can find further details on the generation of the
radiation in appendix~\ref{sec:genISRrad} and appendix~\ref{sec:genFSRrad} for ISR and FSR, respectively.

Since in \POWHEG{} the decomposition in singular regions is driven by the
collinear singularities, in the case of the production of a heavy quark there is
only one FKS singular region, corresponding to ISR from the incoming light
quark. On the other hand, the real matrix element may be enhanced when the extra
gluon is emitted quasi-collinearly to the final-state heavy quark. In this case,
the choice of an ISR hard scale $K_{T}$ is not correct, possibly leading to a
mismodeling of this configurations. A more consistent treatment of the FSR
quasi-collinear region would require its inclusion in the FKS decomposition, see
e.g. ref.~\cite{Buonocore:2017lry}, and the construction of a suitable mapping
for the case of a massive emitter. While different mappings do
exist~\cite{Barze:2012tt,Buonocore:2017lry}, they have been devised in the
context of hadron-hadron collisions. Since there the radiation recoil is shared
by all particles in the final state, the leptonic variables are not preserved in
such mappings, which are then not ideal for DIS processes.

In this work, we follow a different strategy. Despite being potentially large,
the contribution of the quasi-collinear configurations is finite thanks to the
heavy quark mass. Therefore, it can be generated separately as a regular (i.e
non-singular) real component. To this end, we introduce a smooth decomposition
of the real squared amplitude
\begin{equation}
  R = w_{\textrm{QFSR}} R + (1-w_{\textrm{QFSR}})R \equiv R_{\textrm{QFSR}} + R_{\textrm{sing}}\;.
\end{equation}
The contribution $R_{\textrm{sing}}$ contains all soft and/or collinear
singularities and is suppressed in the FSR quasi-collinear configurations, which
is instead dominant in $R_{\textrm{QFSR}}$. Then, we replace
$R \to R_{\textrm{sing}}$ in the \POWHEG{} Sudakov and in the ${\bar B}$
function, while we generate remnant events according to the distribution
\begin{equation}
   R_{\textrm{QFSR}}(\Phi_{R})d\Phi_{R}
\end{equation}
with standard Monte Carlo methods. In order to construct the $w_{\textrm{QFSR}}$
function we introduce the distances of the radiated parton with respect to the
initial-state light quark $d_{\textrm{ISR}}$ and to the final-state massive
quark $d_{\textrm{QFSR}}$
\begin{equation}
  d_{\textrm{ISR}} = \frac{p \cdot k}{p^{ 0}}, \quad d_{\textrm{QFSR}} = \frac{v \cdot k}{v^{ 0}} + m_{v},
\end{equation}
where $p, v$ and $k$ are, respectively, the momentum of the incoming light
quark, of the outgoing heavy quark, with mass $m_v=\sqrt{v^2}$, and of the
radiation, with all energies evaluated in the partonic CM frame. Then, we write
\begin{equation}
  w_{\textrm{QFSR}} = \frac{d_{\textrm{ISR}}}{d_{\textrm{ISR}}+d_{\textrm{QFSR}}}.
\end{equation}
Notice, in particular, that in the soft limit $d_{\textrm{QFSR}} \to m_{v}$
while $d_{\textrm{ISR}} \to 0$, so that $ w_{\textrm{QFSR}}\to 0$, which
ensures that all singularities are contained in $R_{\textrm{sing}}$. The above
remnant mechanism represents our default choice for DIS processes characterised
by a heavy quark in the final state. Furthermore, in order to avoid issues
related to real configurations whose underlying Born gives a vanishing or very
small contribution, we always turn on the \texttt{Bornzerodamp}
mechanism~\cite{Alioli:2010xd} in our \POWHEG{} generator.

The calculation for the heavy quark production processes is performed in the
decoupling scheme with $n_{f}=3$ active flavors in the running of
$\alpha_{S}$ and in the proton. We are dealing with EW processes that are
quark-initiated at the lowest order. As a consequence, the strong coupling is not
renormalised by NLO corrections and the gluon parton density starts to
contribute only at NLO. Therefore, we can consistently adopt PDF sets and
$\alpha_{S}$ running with $n_{f}=4$ without the need of introducing any
correction terms related to the change of scheme, see
e.g.~\cite{Cacciari:1998it}. This matches what is done for all other cases
involving only light quarks, where we consider a $n_{f} = 4$ proton with a
massless charm component to complete the second-generation $SU(2)$ doublet.

We focus on the following representative processes
\begin{itemize}
  \item $e^{ -} +  p \to \nu_{e} + X$(no masses);
  \item $e^{ -} + p \to \nu_{e} + \bar{c} + X$ (production of a massive quark);

  \item $\nu_{\tau} +  p \to \tau^{ -} + X$ (production of a massive lepton),
  \item $\nu_{\tau} + p \to \tau^{ -} + c + X$ (production of a massive quark and a massive lepton),
\end{itemize}
which include all combination of massive quark/lepton in the final state.

In all cases, we consider a reference setup with an incoming lepton with fixed
energy $E_{\ell} = 1\,$TeV scattering off a proton at rest in the laboratory
frame. A cut on the minimum momentum transfer $Q^{ 2}>4\,\textrm{GeV}^{ 2}$ is
applied. The $W$ boson mass is set to $m_{W}=80.419\,$GeV, the $\tau$ mass to
$m_{\tau} = 1.777\,$GeV, the charm mass to $m_c = 1.5\,$GeV, the Fermi constant
to $G_{F} = 1.16639\times 10^{ -5}\,\textrm{GeV}^{-2}$, and the cosine of the
Cabibbo angle to $\cos{\theta_{C}} = 0.97462$. We use the NNPDF3.1 NLO
PDFs~\cite{NNPDF:2017mvq} with $\alpha_{S}=0.118$ and $n_{f}=4$ through the {\tt
  LHAPDF} interface~\cite{Buckley:2014ana}. We adopt a dynamical
scale\footnote{In order to avoid differences due to the scale settings when
  using different mappings, the dynamical scale is computed separately for the
  real and for the underlying Born configurations, turning on the flags {\tt
    btildescalereal} and {\tt btildescalect} in the {\tt powheg.input} file.}
$\mu_{0} =\sqrt{Q^{ 2}+m_{v}^{ 2}}$, where $m_{v}$ is the mass of the
final-state quark. We will refer in the following to
the $Q^{ 2}>4\,\textrm{GeV}^{ 2}$ cut as the definition of our fiducial region.

In table~\ref{tab:xsecs-LO-NLO}, we list the LO and NLO rates in our fiducial
region for
\begin{table}
  \centering
  \begin{tabular}{c|c|c|c}
    process & $\sigma_{\textrm{LO}}$ [pb] ] & $\sigma_{\textrm{NLO}}$ [pb] & $\sigma_{\textrm{NLO}}/\sigma_{\textrm{LO}} - 1 $ [\%]  \\
    \hline
    $e^{ -} + p \to \nu_{e} + X$ & $ 3.7881(3)$ & $3.6741(6)$ & $-3.0\%$\\ 
    $e^{ -} + p \to \nu_{e} + c + X $ & $0.069706(15)$ & $0.1056(4)$ & $+51.5\%$ \\ 
   $e^{ -} + p \to \nu_{e} + c + X,\, m_{c}=0  $ & $0.0644(2)$ & $0.1039(4)$ & $+61\%$ \\
    $\nu_{\tau} + p \to \tau^{ -} + X $ & $4.1228(5)$ & $3.9571(8)$ & $-4.0\%$ \\  
    $\nu_{\tau} + p \to \tau^{ -} + c + X $ & $0.64706(6)$ & $0.6217(2)$ & $-3.9\%$ \\ 
  \end{tabular}
  \caption{\label{tab:xsecs-LO-NLO} Inclusive LO and NLO fiducial rates for the
    four considered processes. The result for a massless charm is also reported
    for comparison. }
\end{table}
producing the final-state lepton (inclusively over any hadronic final
state) or for producing the final-state lepton and the charm. In all
cases but charm electro-production in CC, we find that the NLO
corrections are rather mild, decreasing the rates by a few
percents. The smallness of the NLO corrections is in part a result of
relatively large positive and negative contributions, possibly
occurring among subprocesses in the same $SU(2)$ weak doublet
(neglecting Cabibbo suppressed mixing effects), that cancel to a large
extent in the fiducial rates. We anticipate, therefore, that their
impact on more differential observables may be larger.

On the other hand, in the case of charm electro-production, NLO corrections
increase the LO fiducial rate by $61\%$. It is worth explaining why. The related
process of charm neutrino-production does not feature the same large positive
NLO correction. This is a first indication that the origin of the different
behavior is not related to the massive calculation. Indeed, we can perform the
calculation even for a massless charm by tagging the charm in the final state.
The result for charm electro-production for a massless charm is reported in
table~\ref{tab:xsecs-LO-NLO} for comparison. We found that the NLO correction
has the same pattern as that for the massive charm. This confirms that this
pattern is not associated with mass effects and that logarithms of the charm
mass are not extremely large at the energy scales probed by the considered
lepton-proton scattering process.

On the other hand, a large positive correction can be associated to real
emission processes populating regions of phase space that were inaccessible or
dynamically suppressed at a lower order. Specifically, if we neglect all masses
for simplicity, in the collision $e^-\bar{s} \to \nu_e \bar{c}$, in the partonic
CM, $e^-$ and $\bar{s}$ have opposite helicity and thus the same spin along the
collision direction. In backward scattering also $\nu_e$ and $\bar{c}$ have the
same spin, but opposite to the incoming ones, so that angular momentum
conservation is violated by two units, leading to a $(1-y_{\rm dis})^2$
suppression of the cross section.\footnote{We recall that $\ydis=(1-\cos\theta)/2$,
  where $\theta$ is the scattering angle in the partonic CM frame.} Conversely
in the collision $\nu_e s \to e^- c$ the two incoming particles have opposite
spin, and no such suppression arises.
Therefore, in the case of charm electro-production, real emission processes can
lift the dynamical suppression in the backward scattering region,
leading to a large positive correction to the fiducial rates.

We focus now on the kinematic distributions of the following observables:
\begin{itemize}
  \item the inclusive DIS leptonic variables, $\xbj$ and $\ydis$;
\item the transverse momenta of the leading
  light-flavor (charm-flavor) jet $j_{1}$ ($j^{ c}_{1}$) for the light
  (heavy) quark case and of the second leading light-flavor jet
  $j_{2}$;
\item the rapidity $y_{j_{1}}$($y^{c}_{j_{1}}$) of the
  leading light-flavor (charm-flavor) jet in the lab frame and
  the
  lepton-jet
  $\Delta R_{\ell j_{1}} = \sqrt{ \Delta^{ 2} y_{\ell j_{1}} +
    \Delta^{ 2} \phi_{\ell j_{1}}}$ ($\Delta R_{\ell j^{c}_{1}})$
  separation.
\end{itemize}
We define the jets using the anti-$k_{T}$ clustering
algorithm~\cite{Cacciari:2008gp} as implemented in
\fastjet{}~\cite{Cacciari:2011ma}, with radius $R=1$ and using the $E$-scheme
for the recombination. The jets are required to have a minimum transverse
momentum of $0.1\,$GeV. At parton level, jets containing any charm quarks and/or
anti-quarks are considered charm-flavor jets.

We consider two variants
of the momentum mappings:
\begin{itemize}
  \item ``global'': the ISR mapping preserves the momentum of the incoming
        lepton and the invariant mass of the underlying Born system (simple
        mapping for DIS), the FSR mapping preserves both incoming partons
        momenta (standard FKS FSR mapping implemented in the \POWHEGBOX{});
  \item ``dis'': both ISR and FSR mappings preserve the momenta of the incoming
  and outgoing leptons.
\end{itemize}
In fixed order perturbation theory, the two options must provide equivalent
results. Differences may appear at the level of the events generated according
to the \POWHEG{} formula.

\begin{figure}[htbp]
  \centering
  \includegraphics[width=0.43\textwidth]{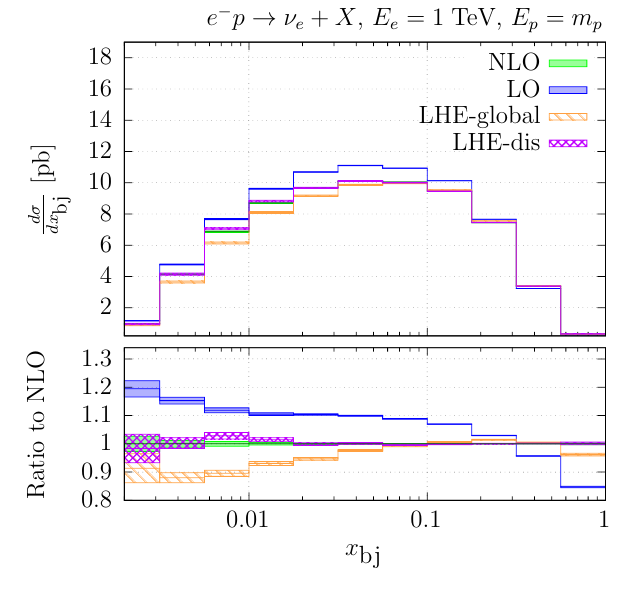} \hspace{0cm} \hfil
  \includegraphics[width=0.43\textwidth]{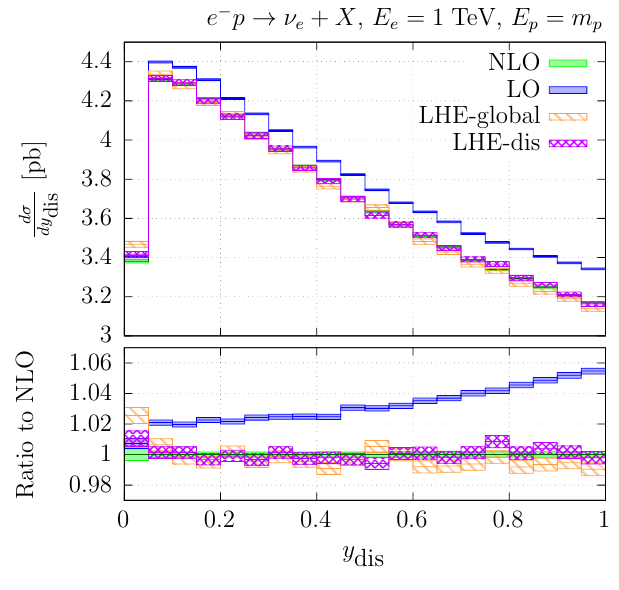}
  \\
  \includegraphics[width=0.43\textwidth]{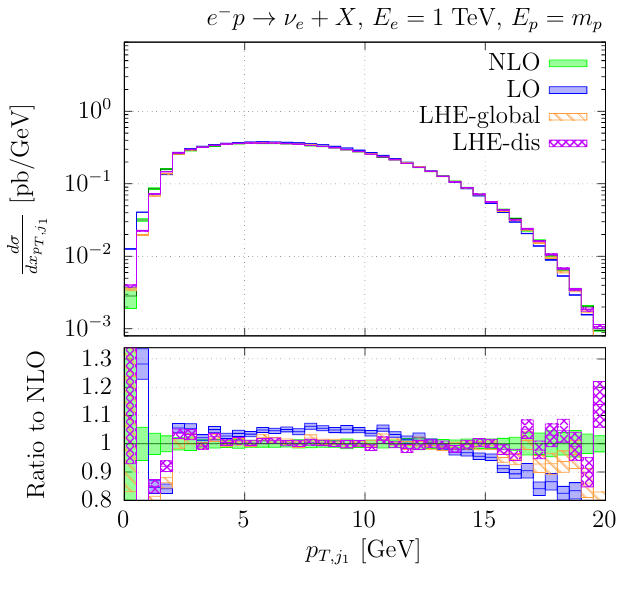} \hspace{0cm} \hfil
  \includegraphics[width=0.43\textwidth]{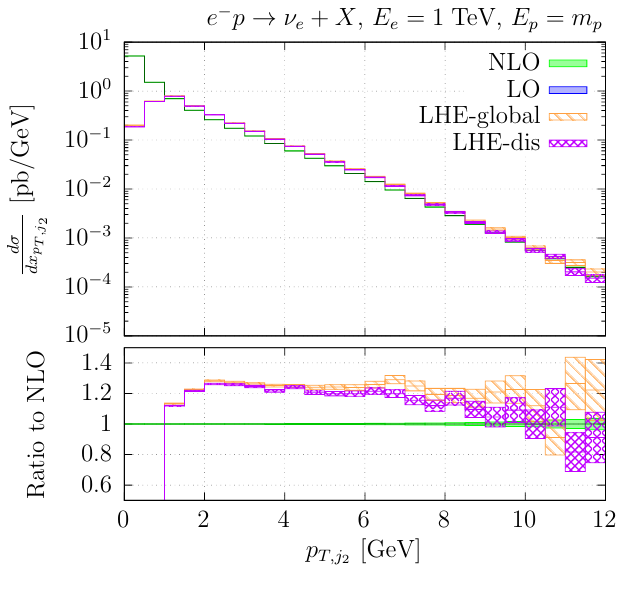}
  \\
  \includegraphics[width=0.43\textwidth]{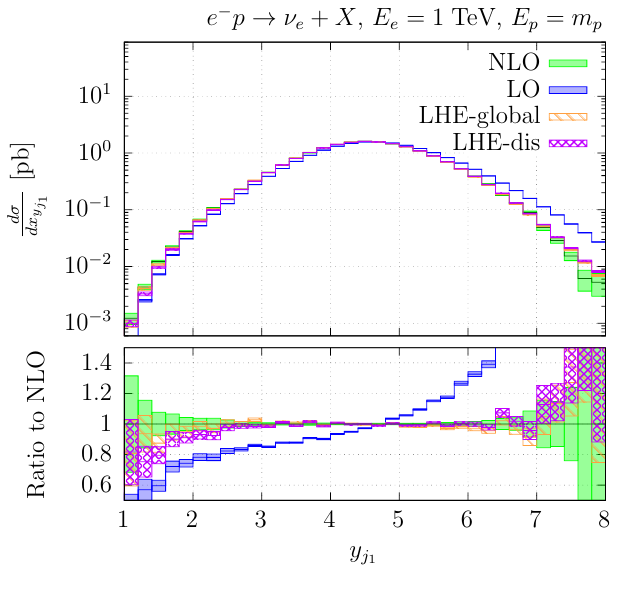}  \hspace{0cm} \hfil
  \includegraphics[width=0.43\textwidth]{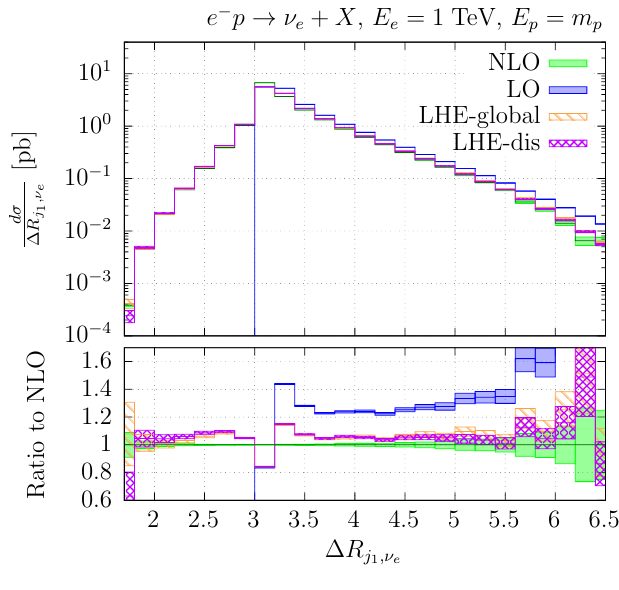}
  \caption{\label{fig:el-CC-0} \footnotesize Collection of differential
    distributions for the scattering of a $1\,\textrm{TeV}$ electron off a
    proton at rest: Björken variable $\xbj$, inelasticity $\ydis$, transverse
    momentum of the leading jet $p_{T,j_{1}}$, transverse momentum of the second
    jet $p_{T,j_{2}}$, rapidity of the leading jet $y_{j_{1}}$, lepton-jet
    separation $\Delta R_{j_{1},\ellp}$. LO predictions are displayed in blue,
    NLO ones in green, results obtained at the \POWHEG{} event level in orange
    and in purple adopting, respectively, the ``global'' (LHE-global) and the
    ``dis'' (LHE-dis) settings. Ratios to NLO predictions are given in the
    bottom panels. }
\end{figure}

\begin{figure}[htbp]
  \centering
  \includegraphics[width=0.457\textwidth]{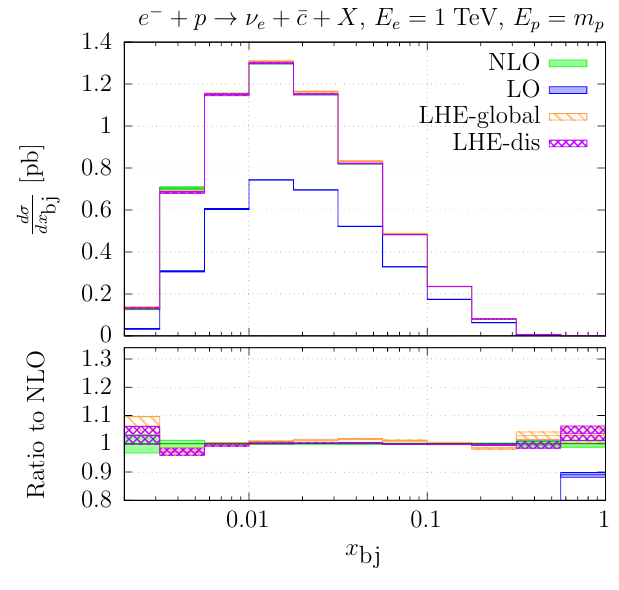} \hspace{0cm} \hfil
  \includegraphics[width=0.457\textwidth]{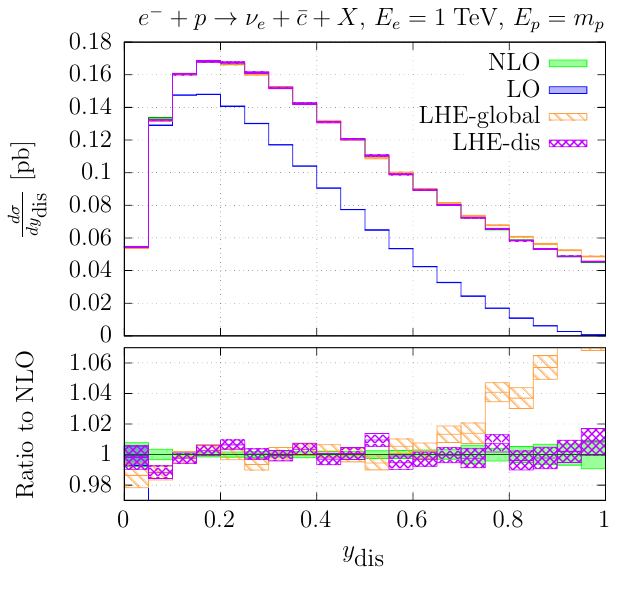}
  \\
  \includegraphics[width=0.457\textwidth]{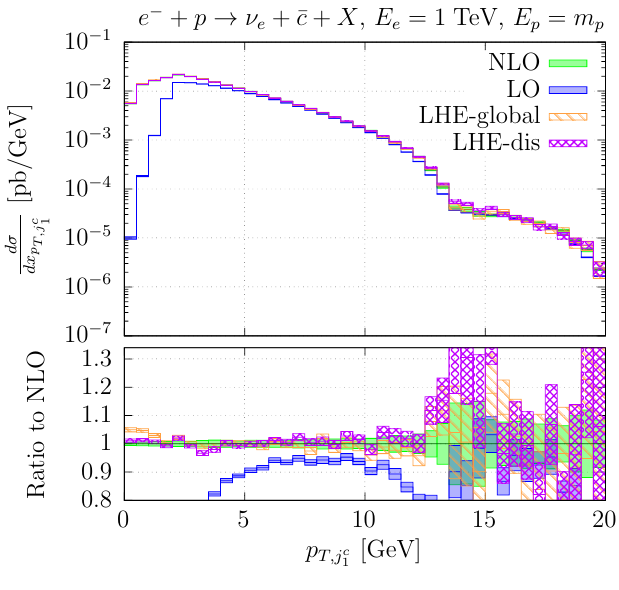} \hspace{0cm} \hfil
  \includegraphics[width=0.457\textwidth]{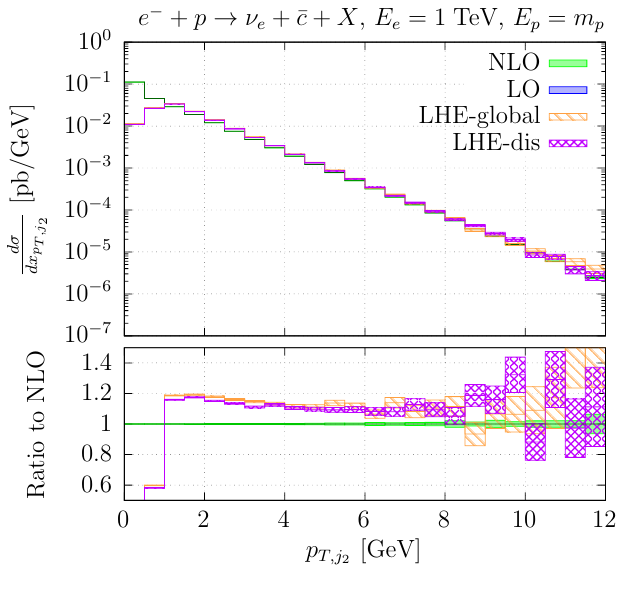}
  \\
  \includegraphics[width=0.457\textwidth]{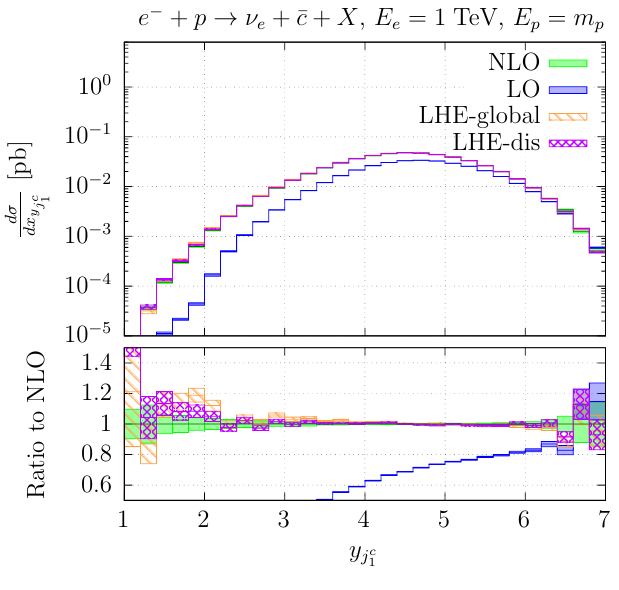}  \hspace{0cm} \hfil
  \includegraphics[width=0.457\textwidth]{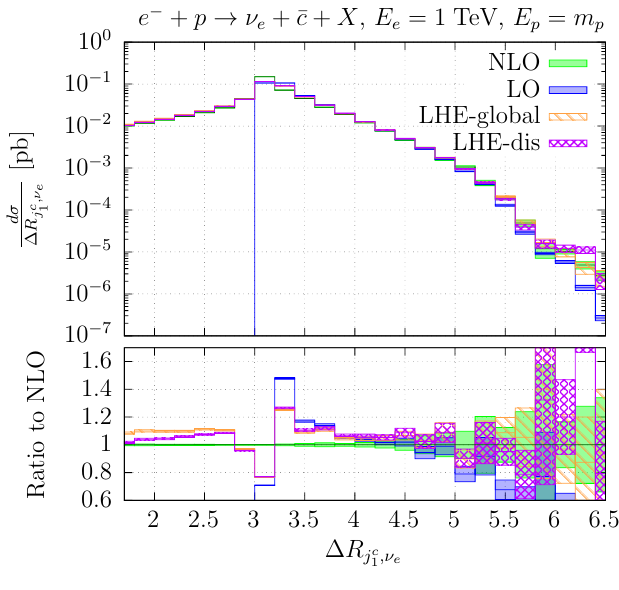}
  \caption{\label{fig:el-CC-1p5} As in Fig~\ref{fig:el-CC-0} for the process
    $e^{-} + p \to \nu_{e} + \bar{c} + X$.
    The leading jet $j_{1}^{c}$ is the leading charm-flavor jet. }
\end{figure}

\begin{figure}[htbp]
  \centering
  \includegraphics[width=0.457\textwidth]{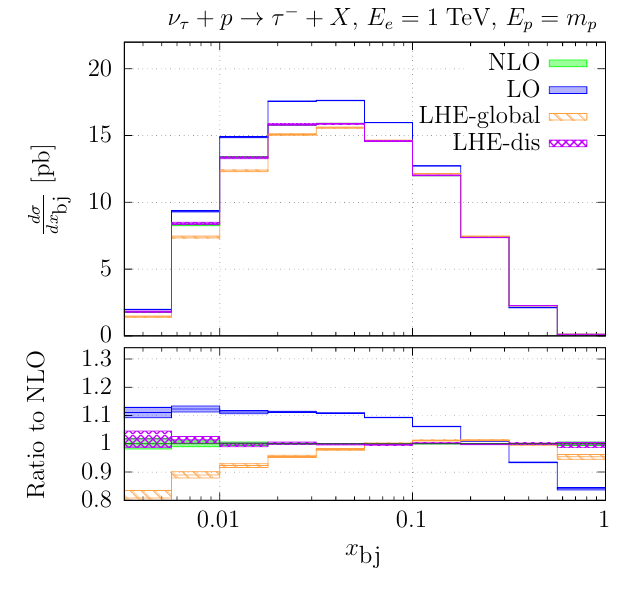} \hspace{0cm} \hfil
  \includegraphics[width=0.457\textwidth]{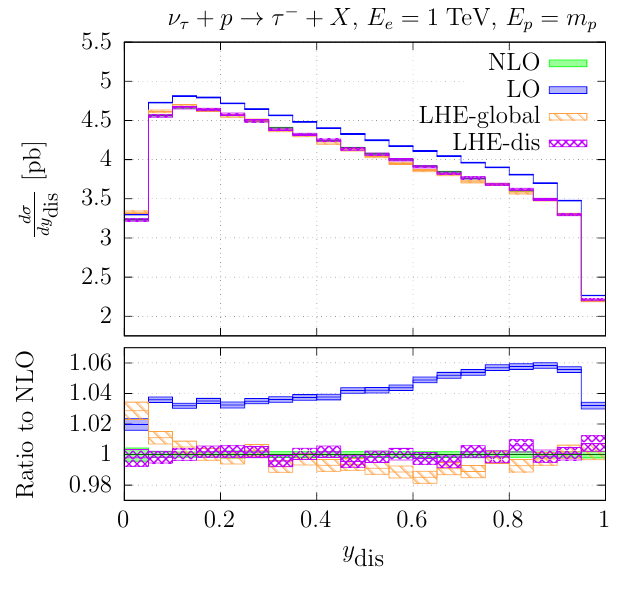}
  \\
  \includegraphics[width=0.457\textwidth]{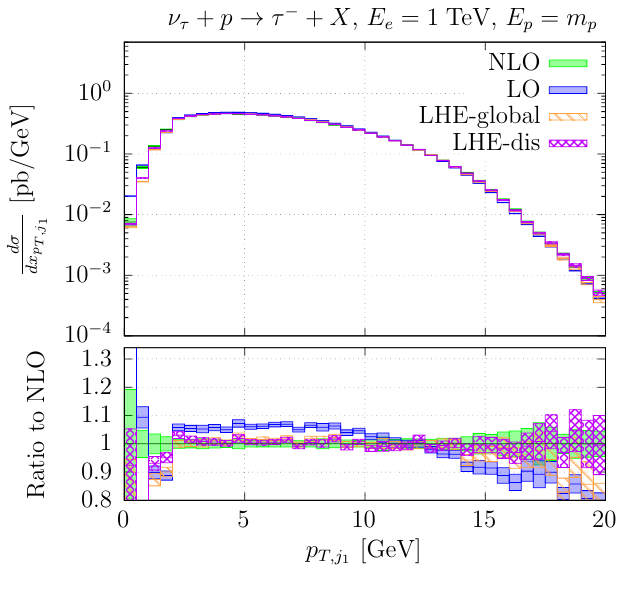} \hspace{0cm} \hfil
  \includegraphics[width=0.457\textwidth]{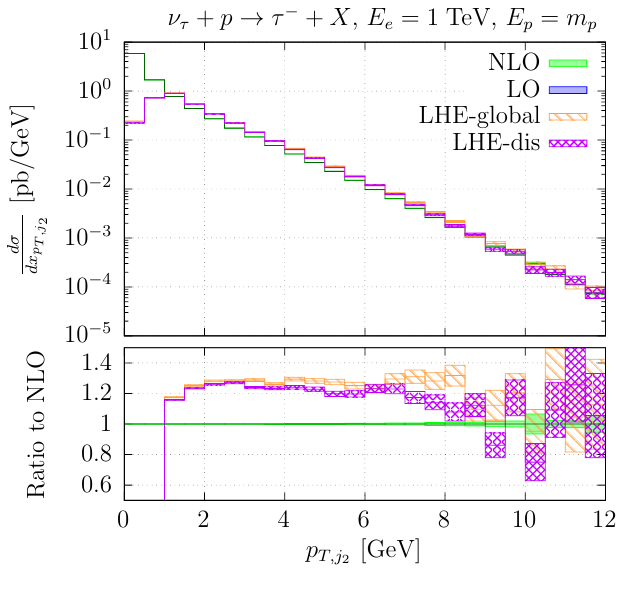}
  \\
  \includegraphics[width=0.457\textwidth]{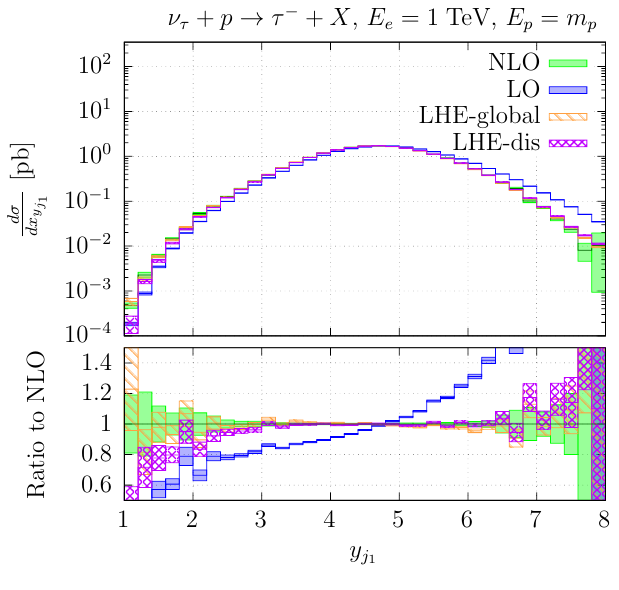}  \hspace{0cm} \hfil
  \includegraphics[width=0.457\textwidth]{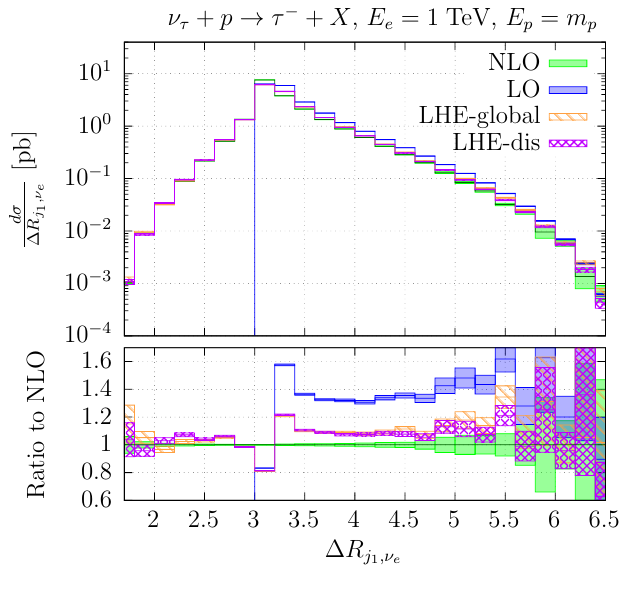}
  \caption{\label{fig:vt-CC-0} As in Fig~\ref{fig:el-CC-0} for the process
    $\nu_{\tau} + p \to \tau^{ -} + X$.}
\end{figure}

\begin{figure}[htbp]
  \centering
  \includegraphics[width=0.457\textwidth]{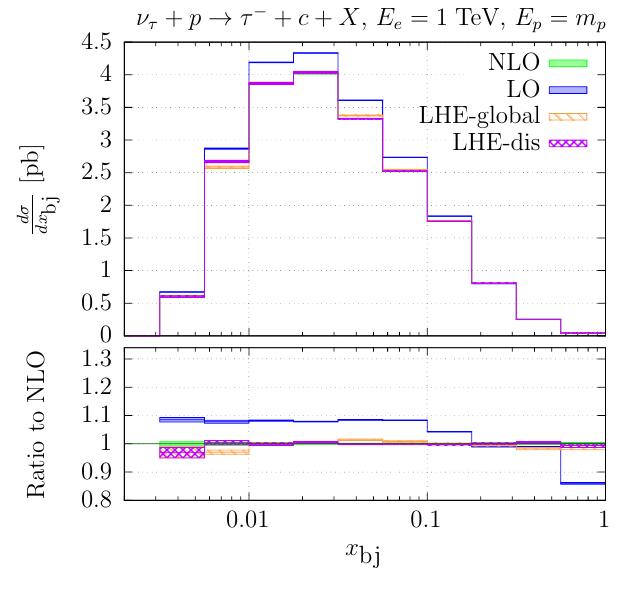} \hspace{0cm} \hfil
  \includegraphics[width=0.457\textwidth]{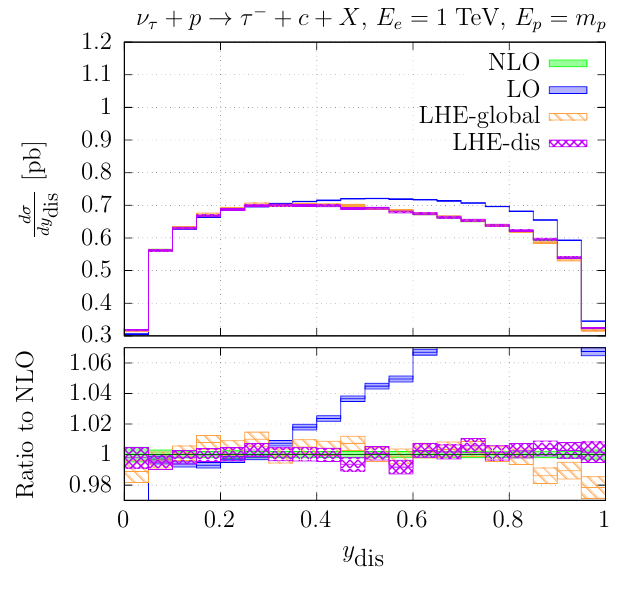}
  \\
  \includegraphics[width=0.457\textwidth]{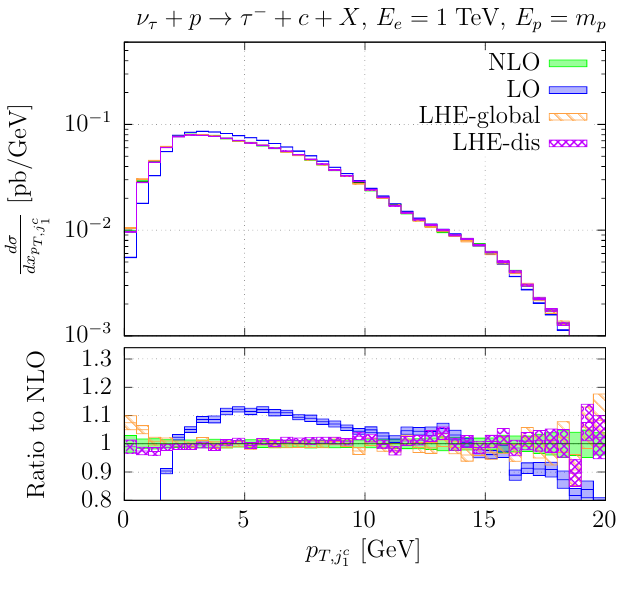} \hspace{0cm} \hfil
  \includegraphics[width=0.457\textwidth]{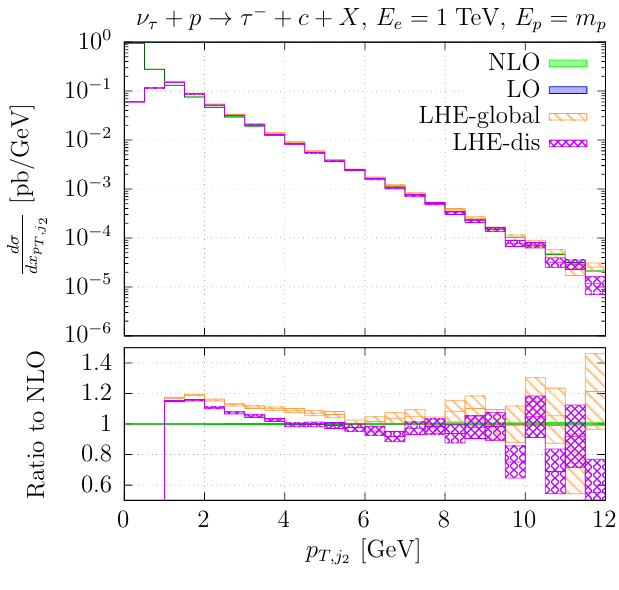}
  \\
  \includegraphics[width=0.457\textwidth]{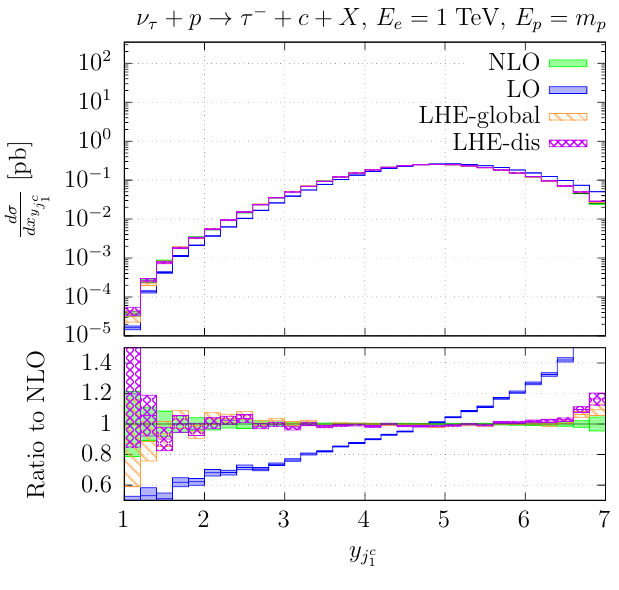}  \hspace{0cm} \hfil
  \includegraphics[width=0.457\textwidth]{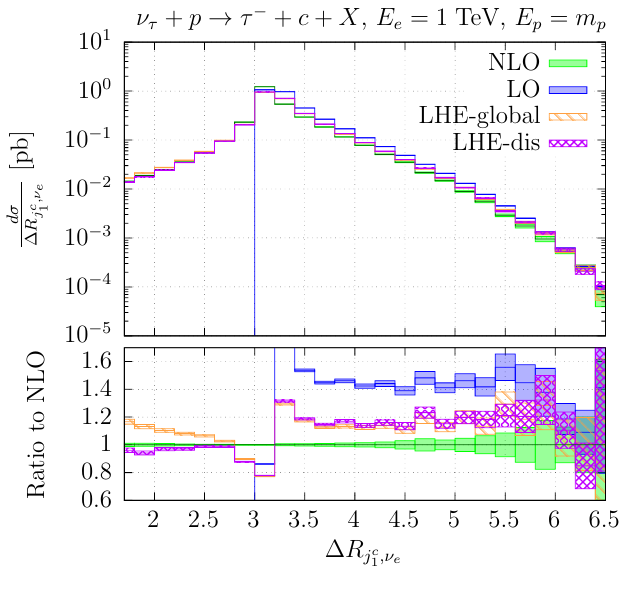}
\caption{\label{fig:vt-CC-1p5}  As in Fig~\ref{fig:el-CC-0} for the process
    $\nu_{\tau} + p \to \tau^{ -} + c + X$. The leading jet $j_{1}^{c}$ is the leading charm-flavor jet.}
\end{figure}

The kinematic distributions are reported in figures~\ref{fig:el-CC-0}-\ref{fig:vt-CC-1p5}. In all plots, we compare predictions obtained
\begin{itemize}
  \item at LO (blue);
  \item at NLO (green);
  \item at the level of the \POWHEG{} events (LHE) with the ``global'' option (orange);
  \item at the level of the \POWHEG{} events with the ``dis'' option (purple).
\end{itemize}
The figures display statistical uncertainties as bands, while scale
uncertainties are not taken into account.

Upon analyzing the first two curves, it is evident that NLO radiative
corrections, while very mild for fiducial rates, can have a significant impact
on more differential observables. Excluding the case of charm
electro-production, it can be observed that NLO corrections have a similar
pattern for all the considered processes. Specifically, the corrections have a
significant impact on the shapes, including the DIS variable $\xbj$, with the
corrections reaching levels of $10-15\%$. The rapidity spectrum of the leading
jet is particularly affected, with NLO corrections of about $50\%$ that shift
the distribution towards smaller rapidities. Correspondingly, the
$\Delta R_{j_{1},\ellp}$ separation between the leading jet and the outgoing
lepton shows a significant NLO correction, with a $30-40\%$ decrease at high
values of separation. At LO, the leading jet and the outgoing lepton were
back-to-back in the transverse plane, resulting in a $\Delta R_{j_{1},\ellp}$
distribution that began at $\pi$. This restriction is lifted at NLO due to the
additional real radiation.

As observed for the fiducial rates, when it comes to charm electro-production,
NLO corrections lead to a significant increase in the normalisation corrections
of approximately $50\%$. This increase can also be seen in the differential
distributions, specifically in the $y_{\rm dis}$ distribution,
  where the $(1-y_{\rm dis})^2$ suppression of the LO result is clearly visible.
  The same suppression also plays a role in the rapidity distribution of the
  leading charm jet, and also in its transverse momentum for
  $p_{T,{j_{1}^{c}}}^2\lessapprox Q^2_{\rm min}$, where forward scattering is
  kinematically suppressed, while backward scattering is suppressed at LO, but
  becomes larger due to NLO corrections.

The transverse momentum distribution of the charm jet exhibits a
distinctive change in slope at approximately
$p_{T,{j_{1}^{c}}} = 13\,\textrm{GeV}$. This effect is due to the different
behavior of the dominant anti-strange and Cabibbo suppressed anti-down channels.
The latter extends to higher $p_{T,{j_{1}^{c}}}$ values, resulting in a more
gradual slope change beyond the kink point.

We will now compare the NLO results with the ones generated using the \POWHEG{}
formula at the event level, using the ``global" (LHE-global) and ``dis" (LHE-dis)
settings for the mappings. It is important to remind that in the case of charm
production, only the ISR region is present.

For the leptonic DIS observables we found, as expected, excellent agreement
between NLO and LHE-dis, within their statistical uncertainties. Deviations of
the LHE-global curves are mostly visible in the $\xbj$ distribution, reaching a
few tens of percent for $\xbj<10^{-2}$. For jet observables, LHE-dis and
LHE-global generally provide similar results, with mild deviations mostly seen
in observables more sensitive to the extra radiation, such as the transverse
momentum of the second jet $p_{T,j_{2}}$ and the $\Delta R_{j_{1},\ellp}$
separation, especially in the region of small separations
($\Delta R_{j_{1},\ellp}<\pi$).

When it comes to charm electro-production, the differences between LHE-global
and LHE-dis results for the $\xbj$ distribution are less noticeable compared to
the massless quark cases. On the other hand, there are still approximately $5\%$
differences at high $\ydis$. The reason behind the former observation, which is
also valid in the case of charm neutrino-production, could be due to the fact
that we are solely comparing the differences between the ISR dis mapping and the
one that preserves the neutrino momentum and the invariant mass of the
underlying Born system, while deviations caused by different mappings for FSR
are expected to be larger.

We would like to briefly discuss the comparisons between the NLO results and the
ones obtained at the event level. We will focus on the LHE-dis results and jet
observables, as the leptonic variables are preserved by this choice of mappings.
For leading order (LO) observables such as the transverse momentum and rapidity
spectra of the leading jet, we observe a good agreement between NLO and LHE-dis
results in the bulk, with some deviations towards small values of the transverse
momentum and rapidity. The transverse momentum spectrum of the second jet, which
is entirely due to real radiation, is divergent at NLO towards small values of
transverse momentum. The LHE-dis result has a characteristic Sudakov
suppression, forming a peak for transverse momenta of
$p_{T,j_{2}}\gtrsim1\,\textrm{GeV}$, where deviations from the fixed order
result are at a level of around $20\%$. Then, the two results match at around
$10\,\textrm{GeV}$. Other significant deviations between the NLO and LHE-dis
results are present in the separation $\Delta R_{j_{1},\ellp}$, near
$\Delta R_{j_{1},\ellp}=\pi$, a region which is sensitive to multiple soft
emissions.

\subsection{Comparison with NNLO}

The option ``dis'' is the natural choice for lepton-hadron scattering processes.
Parton shower programs implement recoil schemes that preserve lepton momenta,
which means that predictions for the leptonic DIS variables will remain the same
even after showering the \POWHEG{} events. However, significant differences
exist in the $\xbj$ variables when using the ``global'' recoil as shown in the
previous section. It is important to determine whether these differences are
within the perturbative scale uncertainties. Additionally, NNLO radiative
corrections may modify the leptonic variables. In the following, we compare the
results obtained with the two mappings ``dis'' and ``global'', including NLO
scale variations, with available NNLO results.

For the massless case, we extend our structure function program to NNLO, thanks
to the implementation of the higher-order proton structure functions in {\sc
  Hoppet}~\cite{Salam:2008qg,hoppetv130}. The cross section for massive quark production in
CC lepton-hadron scattering has been computed up to NNLO in QCD in
ref.~\cite{Berger:2016inr}. The computation of the corresponding coefficient
functions has been reported in ref.~\cite{Gao:2017kkx}. However, these results
are not yet implemented in a publicly available tool. The coefficient functions
for a massive quark have been previously computed at NNLO in the approximation
of large momentum transfer, $Q^{2} \gg m_{Q}^{2}$, in
ref.~\cite{Buza:1996wv,Blumlein:2014fqa}. We rely on the latter for our
comparison, as recently implemented in \yad{}~\cite{Candido:2024rkr}.

\begin{figure}
  \centering
  \includegraphics[width=0.45\textwidth]{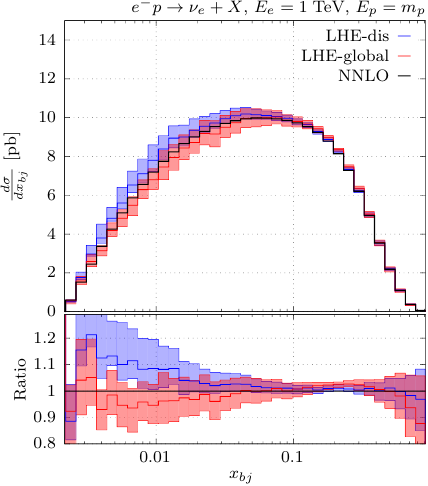}\hfil
  \includegraphics[width=0.46\textwidth]{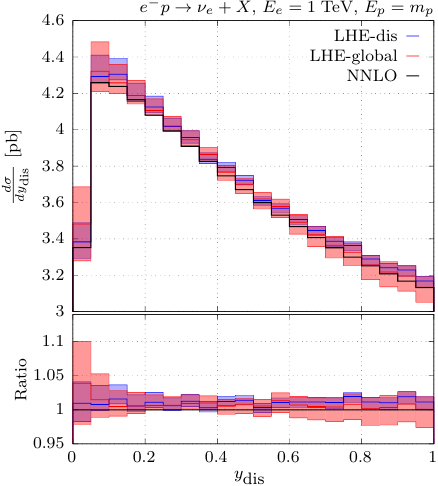}
  \caption{Leptonic kinematic distributions Björken $\xbj$ (left) and $\ydis$
    (right) at the level of the \POWHEG{} events with the ``global''
    (LHE-global) and ``dis'' (LHE-dis) mappings. Scale uncertainties are shown
    as bands. Predictions at NNLO are reported for comparison. Ratios to NNLO
    are shown in the bottom panels.}
\label{fig:cmp_massless_NNLO}
\end{figure}

We focus on the processes $e^{-}+p \to \nu_{e} + X$ and
$e^{-}+p \to \nu_{e} + \bar{c} + X$, beginning with the massless case. The
results for the leptonic variables Björken $\xbj$ and $\ydis$ are displayed in
figure~\ref{fig:cmp_massless_NNLO}.
The bands correspond to scale uncertainties computed by the customary
seven-point scale variation around the reference scale
$\mu^{2}_{F}=\mu^{2}_{R}=Q^{2}$. The Cabibbo angle is set to $\theta_C=0$ for
this comparison. Upon inspection of the $\xbj$ distribution, we observe that for
$\xbj\lesssim 0.01$, there are sizeable differences between the two recoil
options, with their bands (of the order of $15-20\%$) barely overlapping. In
this region, the NNLO prediction lies between the LHE-dis and LHE-global
results. For larger $\xbj$ values the three predictions are closer to each and
the corresponding uncertainty bands are smaller, of the order of $5\%$,
indicating good perturbative convergence in this region. In the case of the
$\ydis$ distribution, except for the first two bins, differences between the two
mappings are very mild, around $1\%$, and are contained within the uncertainty
bands, which are of the order of a few percent. The NNLO corrections are also
mild and flat, reducing the NLO result by a few percent.

\begin{figure}
  \centering
  \includegraphics[width=0.45\textwidth]{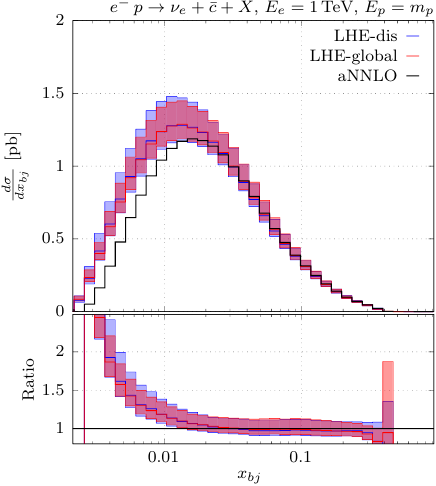}\hfil
  \includegraphics[width=0.45\textwidth]{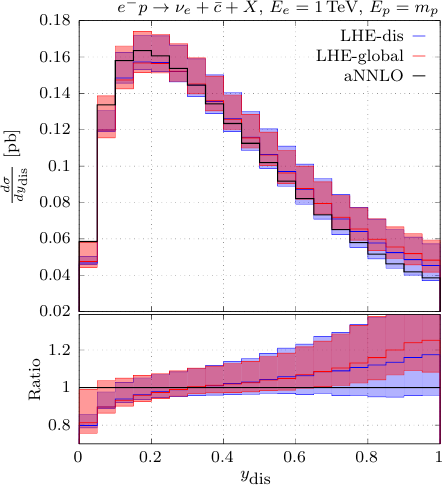}
  \caption{As figure~\ref{fig:cmp_massless_NNLO} for charm electro-production in $e^{-} +p \to \nu_{e} + \bar{c} + X$. Approximate NNLO predictions are obtained with \yad{}\cite{Candido:2024rkr}.}
\label{fig:cmp_massive_aNNLO}
\end{figure}

We turn now to the massive case, shown in figure~\ref{fig:cmp_massive_aNNLO}.
We observe an overall increase in the
perturbative uncertainties and, correspondingly, larger NNLO effects. As observed
and discussed in the previous section, differences between the predictions
obtained with the two mappings are milder than those observed in the massless
case. In the $x \gtrsim 0.01$ region, the approximate NNLO corrections are mild
and flat, indicating good perturbative convergence. However, for $x < 0.01$,
they give a large negative contribution, causing the central aNNLO prediction to
fall outside the NLO uncertainty bands. This region is associated with small
values of the momentum transfer, and the approximation is expected to perform
poorly in this regime. A more meaningful comparison would require the exact NNLO
calculation, which we plan to investigate in future work.

Regarding the $\ydis$ distribution, there is an overall better perturbative
consistency as the aNNLO central prediction is mostly encompassed by the NLO
uncertainties bands. The inclusion of aNNLO corrections brings a non-trivial
shape distortion which leads to a sizeable softening the spectrum. At very high
and very low $\ydis$ values, the aNNLO effects can reach up to $15-20\%$.
However, at very low $\ydis$, the approximation is expected to worsen since this
region probes small values of $Q^{2}$.

\section{Pheno}    
\label{pheno}

In the previous section, we have introduced a new generator for NC and CC DIS
lepton-nucleon processes reaching NLO+PS accuracy. In particular, we have
discussed the impact of NLO corrections and some theoretical aspects related to
the choice of the momentum mappings and their impact on differential
distributions for the generated parton level events, before feeding them to a
SMC program.

We now consider the full simulation chain including showering and hadronisation
effects using \PythiaEight{}~\cite{Sjostrand:2014zea} and adopting the recoil
option {\tt SpaceShower:dipoleRecoil}, which is suitable for DIS
processes.\footnote{We remind that \PythiaEight{} adopts as default a global
  recoil scheme for initial state radiation.} We present a selection of
comparisons with available data from electron(positron)-proton collisions at
HERA as well as predictions for neutrino DIS processes for the ongoing
FASER$\nu$ and SND@LHC, and for the upcoming SHiP experiments at CERN. For the
latter case, we will briefly describe how to deal with a broad band beam of
incoming neutrinos.

In particular, we focus on charm CC DIS production, either at the level of
flavoured charm jet or at particle level of charmed mesons and baryons, and on
the production of tau leptons in neutrino interactions. Charm CC DIS production
is relevant for constraining the strange content of the proton. The current
measurement with incoming charged leptons performed by the ZEUS
collaboration~\cite{ZEUS:2019oro} is affected by large uncertainties, and the
situation is expected to improve with the proposed EIC
experiment~\cite{Arratia:2020azl}. On the other hand, for neutrino beams with an
emulsion detector the identification of charm is topological, and thus has a
very high efficiency and purity.

\subsection{DIS in the forward region at HERA}

In the following, we compare NLO+PS predictions with the single-jet measurements
performed by the ZEUS collaboration for differential distributions in the
laboratory frame~\cite{ZEUS:2005ukc}. The theory uncertainties on the
predictions are obtained from the customary seven-point scale variation of
renormalisation and factorisation scales around a central value of
$\mu^{2}_{F} = \mu^{2}_{R} = Q^{2}$. We adopt the same physical parameters and
pdf set as in Sec.~\ref{sec:validatio-LHE}.

The ZEUS measurement of differential distributions for jet production in the
laboratory frame~\cite{ZEUS:2005ukc} is based on data that were taken colliding
protons with energy of $E_{p} = 820\,\textrm{GeV}$ and positrons with energy of
$E_{e} = 27.5\,\textrm{GeV}$, i.e. at a centre of mass energy of
$\sqrt{s}=300.3\,\textrm{GeV}$. Jets are reconstructed using the $k_{T}$
clustering algorithm in the longitudinally invariant mode ($E_{T}$-weighted
recombination scheme). The experimental analysis studied three regions of
inclusive jet production in phase-space. We focus on the most inclusive region,
called ``global". This region, which is expected to be well-described by the
quark-model picture, is defined by the conditions
\begin{equation}
  Q^{2}>25\,\textrm{GeV}^{2}, \quad \ydis>0.04, \quad E^{\prime}_{e}>10\,\textrm{GeV},
\end{equation}
where $E^{\prime}_{e}$ is the energy of the scattered positron, and at least one
jet satisfying
\begin{equation}
  E_{\textrm{jet}} > 6\,\textrm{GeV},\quad -1 < \eta_{\textrm{jet}} < 3.
\end{equation}

\begin{figure}
  \centering
  \includegraphics[width=0.48\textwidth]{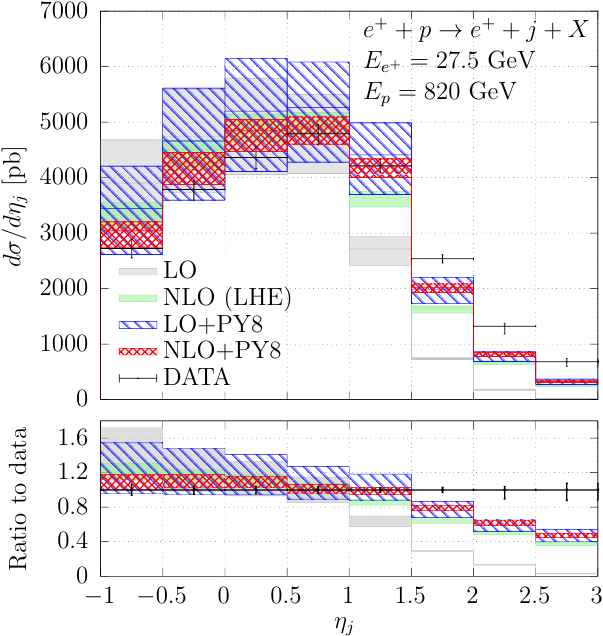}
  \includegraphics[width=0.48\textwidth]{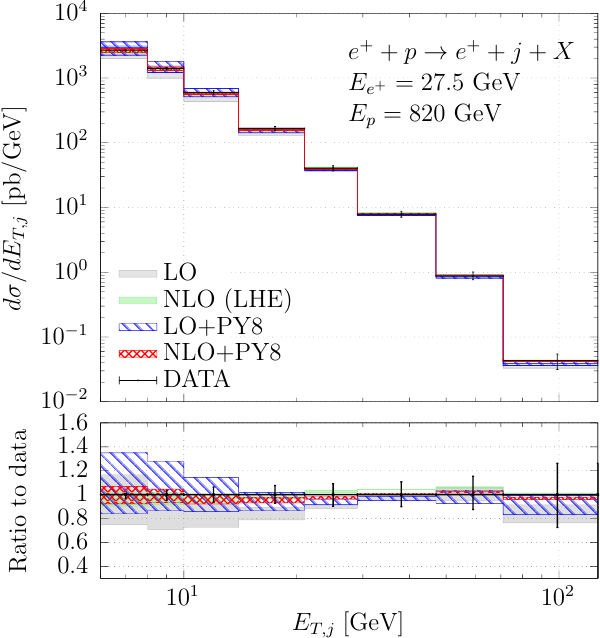} \\
  \includegraphics[width=0.48\textwidth]{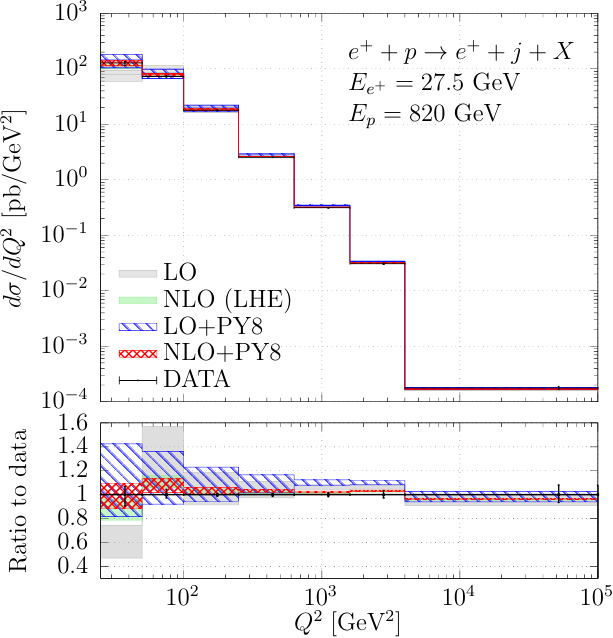}
  \includegraphics[width=0.48\textwidth]{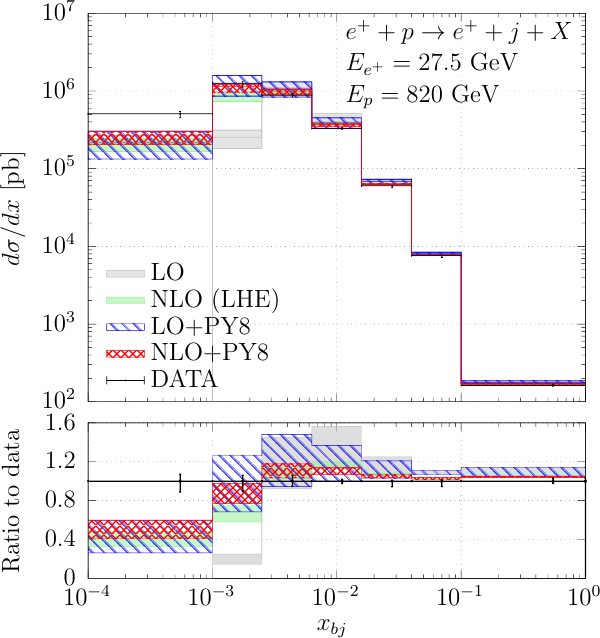}
  \caption{\label{fig:ZEUS} Comparisons of theoretical predictions at LO (in
    gray), NLO (LHE) (in green), LO+PS (in blue) and NLO+PS (in red) with ZEUS
    measurements~\cite{ZEUS:2005ukc} for kinematic distributions in single
    inclusive jet production: the leading jet pseudo-rapidity ($\eta_{j}$) and
    transverse energy ($E_{T,j}$), momentum transfer ($Q^{2}$), and Björken
    variable ($\xbj$). LO+PS and NLO+PS are matched to \PythiaEight{} shower.
    Ratios to data are displayed in the bottom panels.}
\end{figure}

We compare results at LO, NLO(LHE), LO+PS, and NLO+PS with the experimental
measurements for the leading jet pseudo-rapidity ($\eta_{j}$) and transverse
energy ($E_{T,j}$), momentum transfer ($Q^{2}$), and Björken variable ($\xbj$),
as shown in figure~\ref{fig:ZEUS}. For a meaningful comparison with data
provided at the particle level, predictions obtained through matching to the
parton shower are required. Results at LO and NLO(LHE) are displayed for
reference only. At NLO+PS level, we achieve a much-improved description of the
data, with significantly reduced scale uncertainty bands and central values
closer to the experimental data in the regions where we expect an improvement.
Examples of a kinematic domain where a good agreement is not expected are
provided by the lower-order suppressed regions of high jet pseudo rapidities
($\eta_{j}\gtrsim 1.5$), and of the small $\xbj$ values
($\xbj \lesssim 10^{-3}$). In fact, these are still not accurately described even
with the addition of the first extra emission, which is given by the exact
matrix element in the NLO(LHE) and NLO+PS predictions and approximated by the
shower in the LO+PS one. Higher-order corrections are necessary for these
regions. The authors of ref.~\cite{Currie:2018fgr} performed a calculation of
the DIS single-jet inclusive production up to N$^{3}$LO in QCD based on the
projection to Born method~\cite{Cacciari:2015jma} and obtained an excellent
description of the ZEUS data in the ``global'' region. However, their
predictions are at the parton level, and to make a meaningful comparison with
the data, the experimental results have been unfolded from particle to parton
level.

\subsection{Charm CC electro-production at HERA}

The production of charm in CC DIS results in smaller cross sections compared to
NC DIS and photoproduction. This makes measuring charm production more
challenging. However, it is an interesting process because it helps constrain
the strange content of the probed nucleon. The first measurement of charm
production in CC DIS was carried out by the ZEUS
collaboration~\cite{ZEUS:2019oro}, using HERA data in $e^{\pm}p$ collisions at a
centre-of-mass energy of $\sqrt{s}=318\;$GeV. Although the measurement was
affected by large statistical uncertainties, it is an important step towards
better understanding charm production in CC DIS.

The charm cross section has been measured in the kinematic phase space region
defined by the requirements
\begin{equation}\label{eq:fidvol-charmZEUSS}
   200\,\textrm{GeV} ^{2} < Q^{2} < 60000\,\textrm{GeV}^{2}, \quad \ydis<0.9\;,
\end{equation}
into two $Q^{2}$ bins: $200 \,{\rm{GeV}} ^{2} < Q^{2} < 1500\,{\rm{GeV}}^{2}$ and
$1500\, \textrm{GeV}^{2} < Q^{2} < 60000\,\textrm{GeV}^{2}$. Moreover, a region for the
\emph{visible} charm jet has been defined by imposing
\begin{equation}\label{eq:fidvol-charmZEUSS-vis}
  E^{\rm jet}_{T} > 5\, {\rm GeV}, \quad -2.5 < \eta^{} < 2.0
\end{equation}
on the identified charm jet, where the jets are reconstructed with the $k_{T}$
clustering algorithm with a radius parameter $R=1$ in the longitudinally
invariant mode and adopting the $E$-recombination scheme.

In table~\ref{tab:ZeussCharm} we report LO+PS and NLO+PS predictions obtained in
the fixed-flavour-number scheme with $n_{f}=3$ light quarks and a massive charm
for both the charm and the visible charm-jet cross sections in the two $Q^{2}$
bins. We set the charm mass to $m_{c} = 1.28\,$GeV and adopt the
ABMP16.3~\cite{Alekhin:2012ig,Bierenbaum:2009zt} PDF set. Our calculation does
not include a significant contribution due to diagrams with a gluon splitting $g \to c\bar{c}$, which are $\mathcal{O}(\alpha_{s}^{2})$ terms.
These contributions are usually considered as background to the so-called EW
component of the charm production in CC DIS since they are dominated by valence
densities. The EW component, in contrast, is directly sensitive to the strange
density.

Upon inspection of table~\ref{tab:ZeussCharm},
\begin{table}
  \centering
  \footnotesize
  \begin{tabular}{|c|c|c|c|c|c|c|}
    \cline{2-7}
     \multicolumn{1}{c|}{}& \multicolumn{3}{c|}{$\sigma^{e^{+}p}_{c,\rm{vis}}$ [pb]} & \multicolumn{3}{c|}{$\sigma^{e^{+}p}_{c}$ [pb]} \\
    \hline
    $Q^2$ range (GeV$^{2}$) & LO+PS & NLO+PS & ZEUS & LO+PS & NLO+PS & ZEUS \\
    \hline
    $200-1500$ & $2.085_{-6.2\%}^{+4.7\%}$ & $2.501_{-3.7\%}^{+4.8\%}$ & $4.1 \pm 2.0$ & $2.657_{-4.5\%}^{+3.3\%}$ & $4.248_{-4.3\%}^{+5.7\%}$ & $8.7 \pm 4.1$  \\
    $1500-60000$ & $0.9921_{-1.5\%}^{+1.3\%}$ & $1.028_{-5.1\%}^{+6.9\%}$ & $-0.7 \pm 2.0$ & $1.157_{-2.2\%}^{+2.2\%}$ & $1.604_{-6.2\%}^{+8.2\%}$ & $-1.2 \pm 3.9$ \\
    \hline
  \end{tabular}\caption{\label{tab:ZeussCharm} LO+PS and NLO+PS predictions for charm production cross sections in CC DIS $e^{+}p$ at HERA within the kinematic phase space defined in eq.~\eqref{eq:fidvol-charmZEUSS}. The visible cross sections refer to the additional visibility cuts on the charm jet in eq.~\eqref{eq:fidvol-charmZEUSS-vis}. Corresponding cross sections measured by the ZEUS collaboration~\cite{ZEUS:2019oro} are also reported. }
\end{table}
we see that radiative corrections have a more significant impact on the first
$Q^{2}$ bin at lower $Q^{2}$ values and in the more inclusive setup compared to
the visible charm-jet region. For the more inclusive region, they reach up
to $+60\%(+45\%)$ in the first(second) bin in $Q^{2}$, with no overlap between
the LO+PS and NLO+PS results. The slightly larger correction in the first bin is
related to larger values of $\as$ at smaller scales. The NLO corrections turn
out to be smaller when applying the visibility cuts on the charm jet. This is
likely due to the fact that these cuts affect the population of large $\ydis$
region, that for charm electro-production gets larger positive NLO corrections,
as already discussed in the previous section.

Scale uncertainties are obtained by independently
varying the renormalisation $\mu_R$ and factorisation $\mu_F$ scales around the
central reference scale $\mu_0 = \sqrt{Q^2+m_c^2}$ by a factor of two up and
down, with the additional constraint $1/2 < \mu_R/\mu_F < 2$. The results
obtained from LO calculations tend to underestimate the rates and corresponding
theoretical uncertainties. The inclusion of higher-order corrections is,
therefore, necessary.

\subsection{DIS with a neutrino flux}

We start by considering a flux of incoming neutrinos given as a binned histogram
of the neutrino energy in the laboratory frame of the $\nu N$ collision. In
particular, the histogram defines the normalised flux function $f$ as
\begin{equation}
  \label{eqn:neutrino_hist}
  f(E_\nu^{\rm{lab}})= \frac{1}{N}\frac{\mathrm{d} N}{\mathrm{d} E_\nu^{\rm{lab}}}.
\end{equation}
Defining $E_\nu^{\rm{max,lab}}$ as the energy of the most energetic neutrino in
the flux, each neutrino involved in the scattering process can be thought as
carrying a fraction $x=E_\nu^{\rm{lab}}/E_\nu^{\rm{max,lab}}$ of the maximal
energy of the beam. Furthermore we set as reference frame the CM of $\nu$ with
energy $E_{\nu}^{\rm{max,lab}}$ and the nucleon at rest. We then define
\begin{equation}
S_H=M^2+2 M E_\nu^{\rm{max,lab}},
\end{equation}
with $M$ denoting the mass of the nucleon. The maximum neutrino energy in the
reference frame is then
\begin{equation}
  \label{eqn:enumaxcms}
  E_\nu^{\rm{max,CM}}=\frac{S_H-M^2}{2\sqrt{S_H}}.
\end{equation}
We then define a neutrino beam density function (BDF) as
\begin{equation}
  \label{eqn:nupdf}
  f_\nu(x) \equiv \frac{1}{N}\frac{\mathrm{d} N}{\mathrm{d} E_\nu^{\rm{lab}}} E_\nu^{\rm{max,lab}}=
   \frac{1}{N} \frac{\mathrm{d} N}{\mathrm{d} x}.
\end{equation}
This is now the boost invariant neutrino BDF we need. Since SMC programs normally
deal with a fixed-energy lepton-nucleon interaction, some extra care is needed
when interfacing them with \POWHEG{}.

\subsection{Setup}
We consider three case studies: the SND@LHC, SHiP and FASER$\nu$ experiments.
For all these
applications with neutrino fluxes we have used the setup of
section~\ref{sec:validatio-LHE}, except that the cut in $Q^2$ has been
set to $Q^2>2\,$GeV$^2$.
The standard seven-point scale variation will also be shown in all plots.
Furthermore, having all the three experiments a
tungsten target, we separately computed the cross section for scattering off
protons and neutrons and combined the results to build the cross section per
tungsten nucleon, that will be denoted as $n_t$ in the plots.

\subsection{$\tau$ neutrinos at SND@LHC}
\label{subsec:neutrinos_SND}
\begin{figure}
  \includegraphics[width=0.5\textwidth]{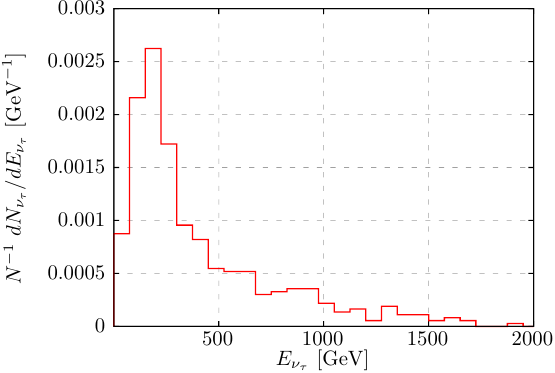}
  \hspace{0.5cm}
  \includegraphics[width=0.5\textwidth]{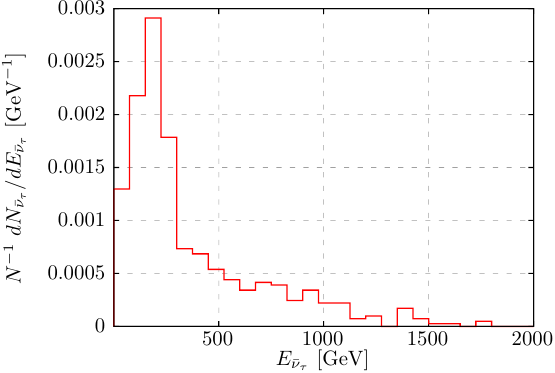}
  \caption{Normalized energy flux of $\nu_{\tau}$ (left panel) and
    ${\bar{\nu}_ {\tau}}$ (right panel) entering the SND@LHC target.}
\label{fig:fluxes-SND}
\end{figure}

As an illustrative example of the computation of fully differential deep
inelastic scattering cross sections initiated by a flux of neutrinos with
variable energy we consider the charged current $\tau$ neutrino and
anti-neutrino interactions at SND@LHC. We have used the normalized flux
simulated by the SND@LHC collaboration~\cite{SNDLHC:2022ihg}, which we show in
figure~\ref{fig:fluxes-SND}.
Here and in the following we assume no errors on the fluxes, that are thus given
as step functions.

\begin{figure}
  \includegraphics[width=0.5\textwidth]{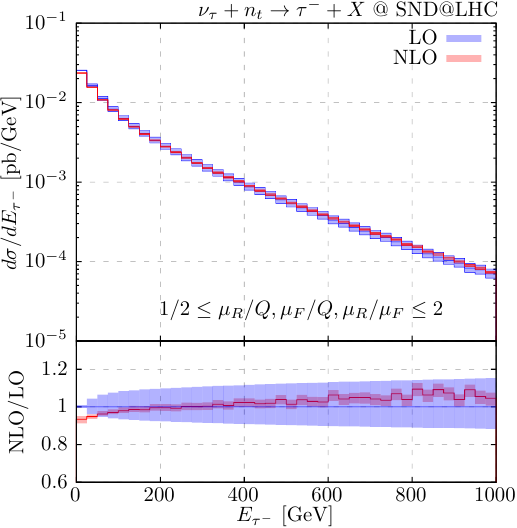}
  \includegraphics[width=0.5\textwidth]{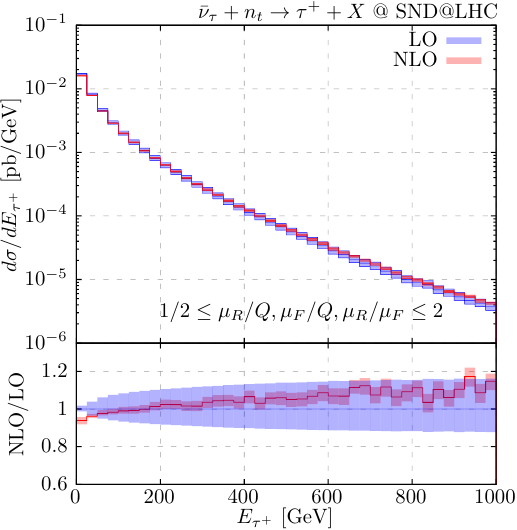}
  \caption{Energy distribution of $\tau^-(\tau^ +)$ produced in charged current
    $\nu_{\tau}~(\bar{\nu}_\tau)$ scattering at the SND@LHC target per tungsten nucleon.}
\label{fig:CCtau-SND}
\end{figure}

In figure~\ref{fig:CCtau-SND} we show the cross sections
per tungsten nucleon as a function of the energy of the produced $\tau$
lepton. NLO corrections are negative and about -5\% in the
dominant small energy region, and mildly rise with energy
reaching +5\% (+10\%) for 1\,TeV $\tau$ (${\bar \tau}$) lepton.
The scale uncertainties are strongly reduced but in the very first bin where
the NLO results lay outside of the LO bands.

\subsection{$\tau$ neutrinos  at SHiP}
\label{subsec:neutrinos_SHiP}

\begin{figure}
  \includegraphics[width=0.5\textwidth]{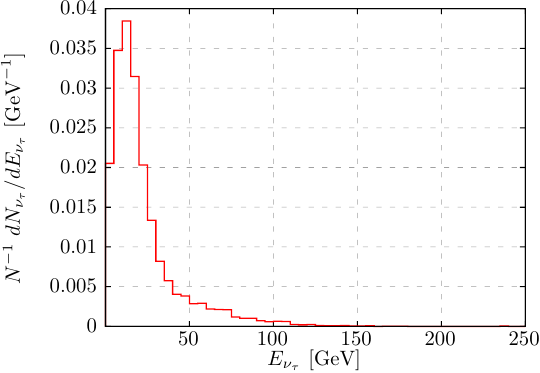}
  \hspace{0.5cm}
  \includegraphics[width=0.5\textwidth]{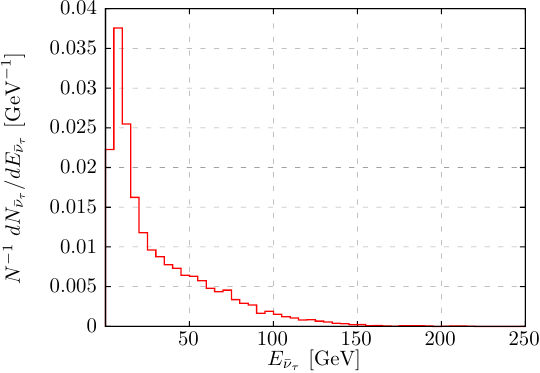}
  \caption{Normalized energy flux of $\nu_{\tau}$ (left panel) and
    ${\bar{\nu}_ {\tau}}$ (right panel) entering the SHiP target.}
\label{fig:fluxes-SHiP}
\end{figure}

\begin{figure}
  \includegraphics[width=0.5\textwidth]{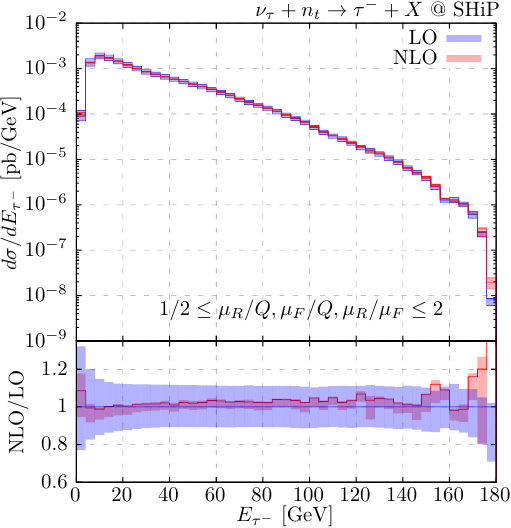}
  \includegraphics[width=0.5\textwidth]{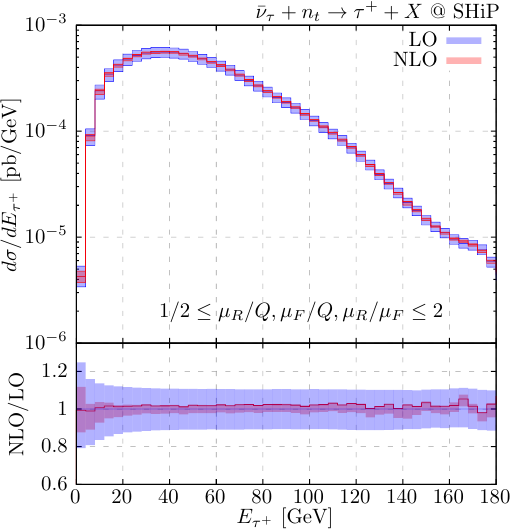}
  \caption{Energy distribution of $\tau^-~(\tau^ +)$ produced in charged
    current $\nu_{\tau}~(\bar{\nu}_\tau)$ scattering at the SHiP target per nucleon of
    Tungsten.
  }
\label{fig:SHiP_Etau}
\end{figure}

We evaluated the DIS cross section induced by an incoming flux of $\tau$
neutrinos and anti-neutrinos interacting in the SHiP target. The fluxes shown in
figure~\ref{fig:fluxes-SHiP} have been generated by the SHiP
collaboration~\cite{SHiP:2015vad}. In figure~\ref{fig:SHiP_Etau} we show the cross
\begin{figure}
  \includegraphics[width=0.5\textwidth]{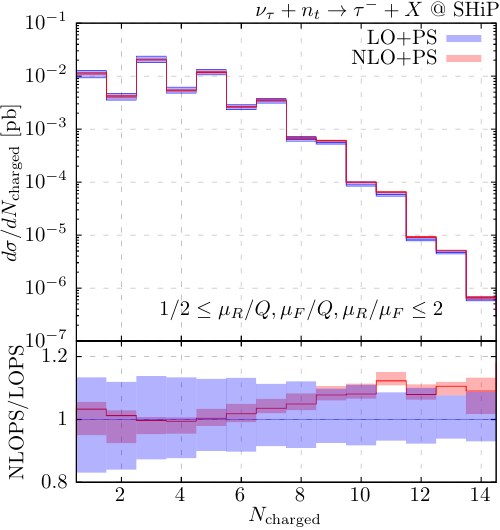}
  \includegraphics[width=0.5\textwidth]{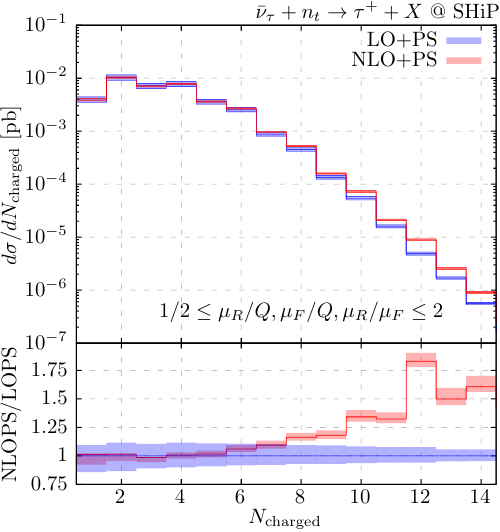}
  \caption{Charged particle multiplicity  produced in charged
    current $\tau$ (anti-)neutrino scattering at the SHiP target per nucleon of
    tungsten.
  }
\label{fig:SHiP_N_Charged}
\end{figure}
section per nucleon as a function of the energy of the produced charged lepton.
Here (and in general also in the following plots)
the radiative corrections reduce considerably the uncertainty band,
becoming non negligible only for very small and very large energies.

By matching our fixed-order computation with \PythiaEight{} we evaluated several
kinematic distributions of variables used to describe the hadronic final state.
We start by showing cross sections as function of the number of charged
particles in the final state after the shower and hadronisation process. We observe that the NLO corrections
tend to increase the multiplicities of charged particles, especially
for the incoming anti-neutrino.\footnote{The larger size of the inclusive cross section
  in the anti-neutrino case has the same explanation that we gave for the
  charm electro-production case, since also in this case we have
  a high-$\ydis$ suppression (at the Born level) of the right-on-left
  collision for valence quarks.}
We also observe the alternating behaviour of the cross section for even-odd
multiplicities. This is explained by the fact that for proton target the multiplicity must
be even (for charge conservation), while for neutron target it must be odd. The tungsten
nucleon is a mixture of the two, with a large prevalence of the neutron. In neutrino
scattering the target down quarks prevail by far, because we have more neutrons
and the neutrons have more down quarks. In the anti-neutrino scattering off protons
an up quark must be hit, thus yielding a cross section larger by about a factor of
two with respect to the neutron case, nearly compensating the larger neutron fraction.
Thus the strong prevalence of the odd multiplicity for neutrinos, and, as can be seen
in the figure, a very
slight prevalence of even multiplicities for anti-neutrinos.

\begin{figure}
  \includegraphics[width=0.5\textwidth]{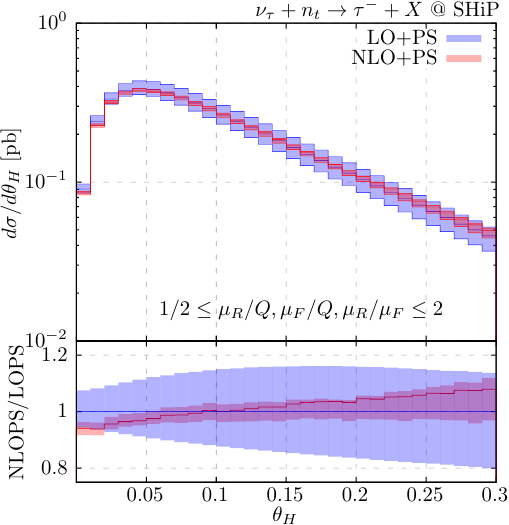}
  \includegraphics[width=0.5\textwidth]{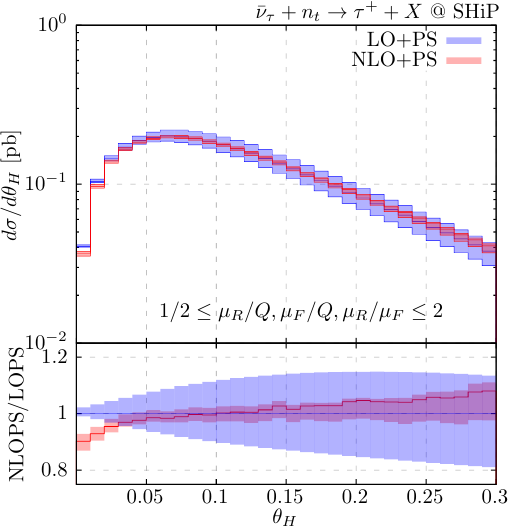}
  \caption{Scattering angle of the hadronic final-state system produced in
    charged current $\nu_\tau$ ($\bar{\nu}_\tau$) scattering at SHiP target per
    nucleon of tungsten. }
\label{fig:SHiP_Hadr_theta}
\end{figure}

In fig.~\ref{fig:SHiP_Hadr_theta} we show the cross section as a
function of the scattering angle $\theta$ of the hadronic final system
in the laboratory frame, for both $\tau$ neutrino and anti-neutrino
scattering. Here we observe a strong
reduction of the scale uncertainty but also regions where the NLO+PS
band is not contained in the LO+PS one. For an incoming neutrino the
radiative corrections have an effect between -5\% up to 10\%.
A similar pattern can be
observed for an incoming anti-neutrino, except for the larger negative
NLO corrections at small angles.

\begin{figure}
  \includegraphics[width=0.5\textwidth]{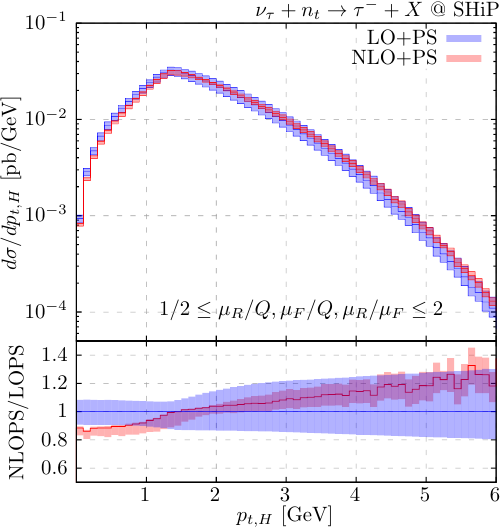}
  \includegraphics[width=0.5\textwidth]{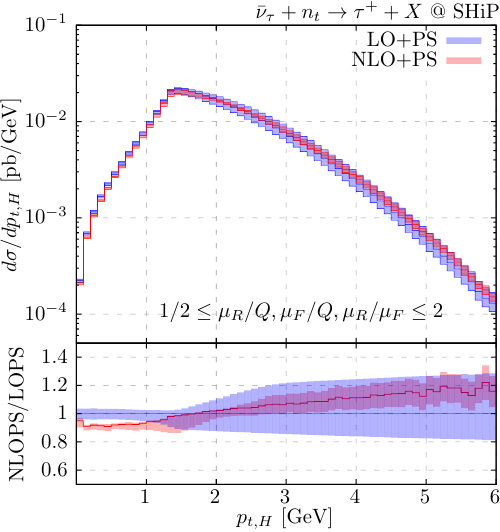}
  \caption{As in fig.~\ref{fig:SHiP_Hadr_theta} for the transverse
    momentum.
  }
\label{fig:SHiP_Hadr_pt}
\end{figure}

In fig.~\ref{fig:SHiP_Hadr_pt} we show the cross section as a function
of the transverse momentum of the hadronic final state, either for an
incoming tau neutrino flux (left panel), or for an anti-neutrino flux
(right panel). We see a similar pattern in both cases.
Looking at $p_t>1\,$GeV we observe that the radiative
corrections are relatively small and contained in the LO+PS
band. Below the cusp, at roughly $p_t=\sqrt{2}\,$GeV, the cross sections
are affected by the $Q^2$ cut, and show a different pattern.
We also see a noticeable change of shape induced by the NLO corrections.

\begin{figure}
  \includegraphics[width=0.5\textwidth]{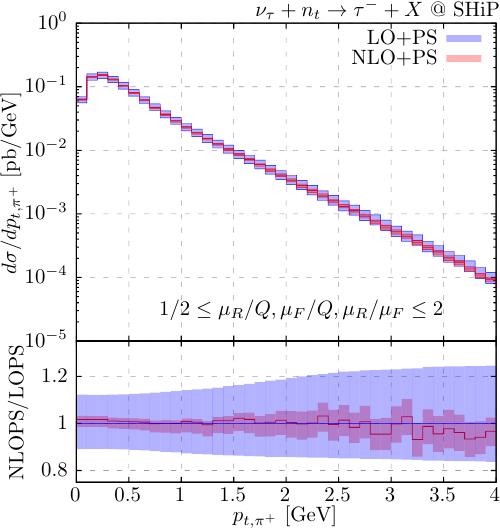}
  \includegraphics[width=0.5\textwidth]{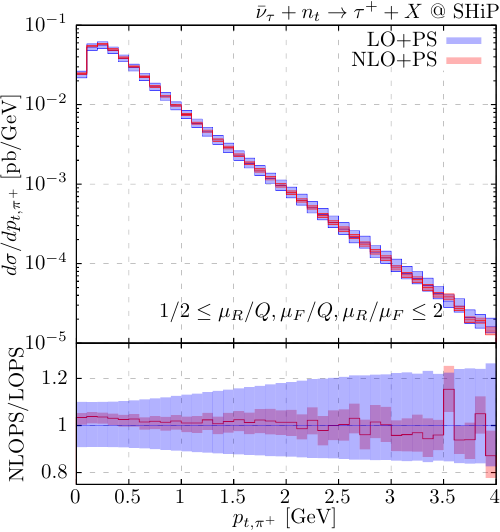}
  \caption{As in fig.~\ref{fig:SHiP_Hadr_pt} for the transverse
    momentum of produced $\pi^+$ for an incoming $\nu_\tau$ ($\bar{\nu}_\tau$)
    flux.
  }
\label{fig:SHiP_pi_pt}
\end{figure}

Finally, in fig.~\ref{fig:SHiP_pi_pt} we show the cross section as a
function of the transverse momentum of inclusively produced
positively charged pions. NLO corrections are mild in the whole range,
with the NLO bands contained in the LO ones in the whole $p_t$
range that we have explored.

\subsection{Charm production at FASER$\nu$}

\begin{figure}
  \includegraphics[width=0.5\textwidth]{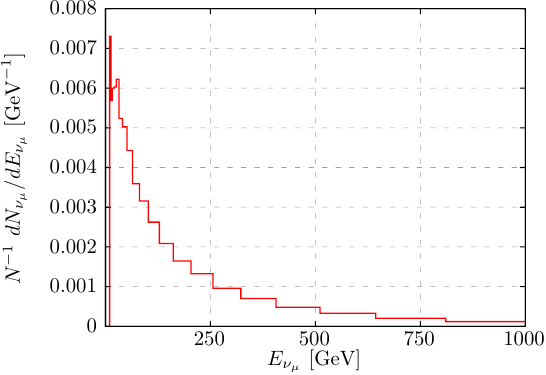}
  \hspace{0.5cm}
  \includegraphics[width=0.5\textwidth]{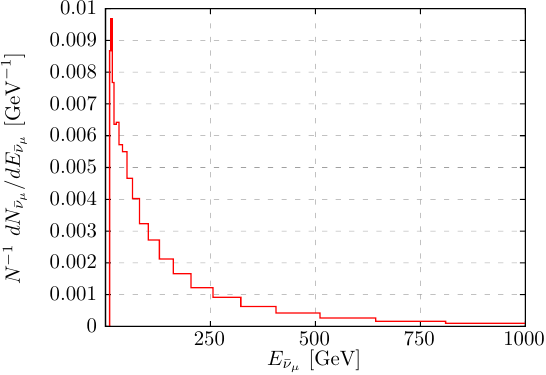}
  \caption{Number of $\nu_{\mu}$ (left panel) and ${\bar{\nu}_
      {\mu}}$ (right panel) entering the FASER$\nu$ target per bin of $\nu$ energy in the
    laboratory frame.}
\label{fig:fluxes-FASER}
\end{figure}

\begin{figure}
  \includegraphics[width=0.48\textwidth]{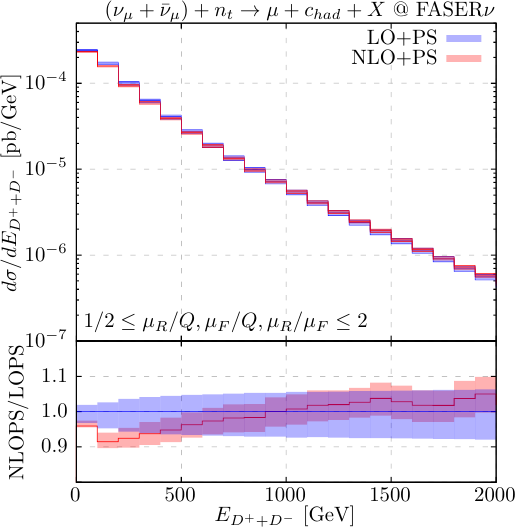}
  \hspace{0.5cm}
  \includegraphics[width=0.48\textwidth]{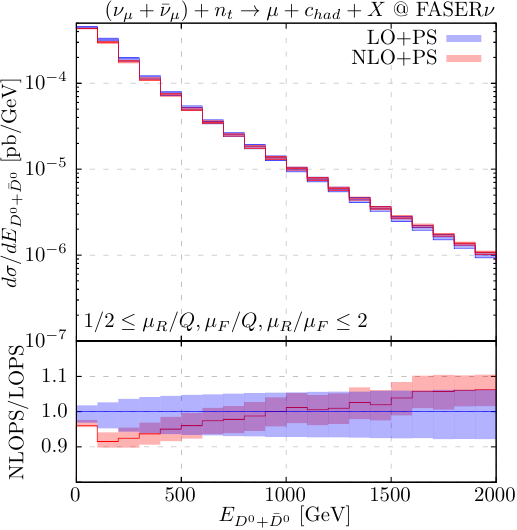}
  \caption{Energy distribution of charged (left) and neutral (right)
    $D$ mesons produced via charged current
    $\nu_\mu+{\bar \nu}_\mu$ events in FASER$\nu$.}
\label{fig:CC-charm-FASER-1}
\end{figure}

\begin{figure}
  \includegraphics[width=0.48\textwidth]{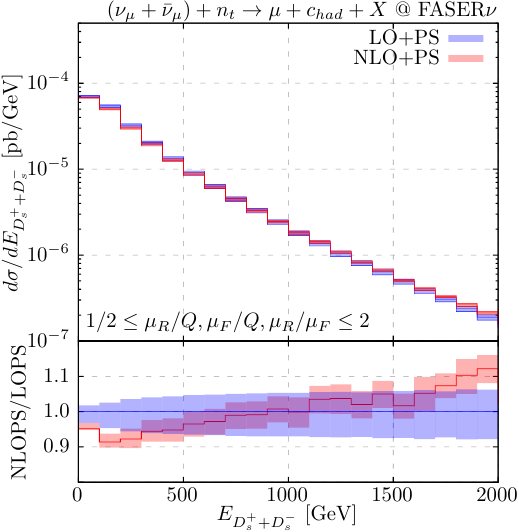}
  \hspace{0.5cm}
  \includegraphics[width=0.48\textwidth]{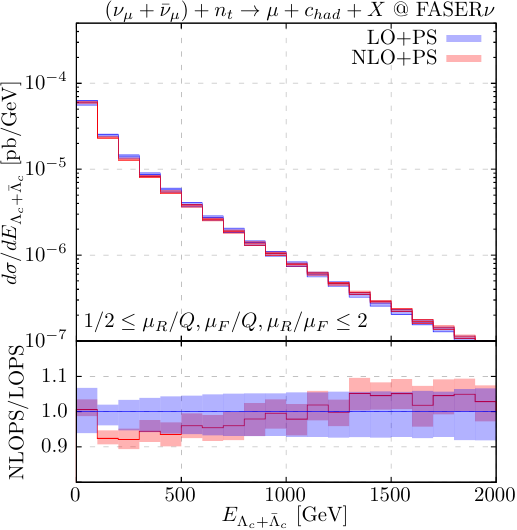}
  \caption{Energy distribution of $D_s$ (left) and $\Lambda_c$ (right)
    particles and antiparticles
    produced via charged current
    $\nu_\mu+{\bar \nu}_\mu$ events in FASER$\nu$.}
\label{fig:CC-charm-FASER-2}
\end{figure}

We have computed the cross section for charm production in muon
neutrino and anti-neutrino CC scattering at FASER$\nu$ using the
fluxes predicted in~\cite{Buonocore:2023kna,FASER:2024ykc} and shown in
fig.~\ref{fig:fluxes-FASER}.
In figures~\ref{fig:CC-charm-FASER-1} and~\ref{fig:CC-charm-FASER-2} the energy
distributions of charmed mesons and $\Lambda_c$ are shown.
Generally speaking, we see that the NLO corrections lead to a hardening
of all distributions.

\section{Conclusions}
\label{conclusion}
In this paper we have presented a new \POWHEG{} event generator to describe deep
inelastic lepton hadron scattering at NLO+PS accuracy. Our code produces results
for charged current as well as neutral current processes initiated by a massless
lepton. The final state lepton and quark can be massless or massive. To build
our code we have developed and implemented new FKS phase space mappings for
initial and final state radiation, that preserve the leptonic variables and are
suitable for the case of massive particles in the final state. These mappings,
described in detail in the appendices, smoothly adapt when the final state
particles are massless.

We have validated our NLO computation through a tuned comparison with an in
house code based upon the DIS lepton hadron cross section expressed in terms of
form factors, which we computed convoluting PDFs with the NLO coefficient
functions.

We have studied the impact of the NLO corrections for the different possible
final states, and the impact of the matching procedure focusing especially on
the (sub-leading) effects related to the choice of the momentum mapping.

We also compared the prediction of our code for the lepton distributions with
available higher order corrections. In order to do this we extended our in house
code by including NNLO form factors available in {\sc Hoppet} for the case of a
massless final state quark, and, for a massive quark, using the approximate
NNLO result implemented in \yad{}~\cite{Candido:2024rkr}.

Having implemented and validated our code, we have shown some illustrative
applications. We first compared our results with two analyses at HERA, one for
charged current $ep$ scattering, and one focusing on single charm production.
Then we moved to incoming neutrino beams, showing predictions for charm and tau
lepton yields at the ongoing experiments FASER$\nu$ and SND@LHC, and at the
upcoming SHiP experiment. In order to run with broad band beams of neutrinos in
the initial state one has to provide a binned histogram with the neutrino flux,
that we have taken from available studies.

Other studies can be easily implemented for past experiments, like the study of
associated charm production in NC $ep$ scattering at HERA, or future DIS studies
at the proposed EIC experiment. Furthermore, our tool may be used in the context
of tau neutrino appearance in atmospheric neutrino oscillations, see
e.g.~\cite{IceCube:2019dqi}, or to include mass effects in top CC DIS production
with very high energy neutrinos from cosmic rays~\cite{Cooper-Sarkar:2011jtt,Garcia:2020jwr}.

Our tool will be soon made publicly available in
the \POWHEGBOX{} repository.\footnote{We remind the reader that implementations for massless unpolarised~\cite{Banfi:2023mhz} and polarised~\cite{Borsa:2024rmh} DIS are available in
the \POWHEGBOX{} repository.}

We conclude by noticing that, being based on the \POWHEGBOX{} framework, our
work can be extended in several directions. NLO electroweak corrections can be
included, and also the NLO QCD corrections can be extended with one extra
radiated parton in the final state. This would allow to promote the computation
to NNLO+PS via a MiNNLO~\cite{Hamilton:2012rf,Hamilton:2013fea,Monni:2019whf}
procedure.

\section*{Acknowledgments}
We are grateful to A. De Crescenzo, F. Kling and M. Fieg for providing us with
the neutrino fluxes at the SND@LHC, SHiP and FASER$\nu$ experiments, A. Candido
and R. Tanjona for support in using \yad, I. Helenius for guidance on the use of
\Pythia{} in the case of a flux of incoming leptons with variable energy. We
thank S.~Ferrario Ravasio, R.~Gauld, B.~J\"ager, A.~Karlberg, P.~Torrielli and
G.~Zanderighi for useful comments on the manuscript. The work of GL has been
partly supported by Compagnia di San Paolo through grant
TORP\_S1921\_EX-POST\_21\_01. The work of PN has been partly supported by the
Humboldt Foundation. The work of LB is funded by the European Union (ERC, grant
agreement No.\,101044599, JANUS). Views and opinions expressed are however those
of the authors only and do not necessarily reflect those of the European Union
or the European Research Council Executive Agency. Neither the European Union
nor the granting authority can be held responsible for them.

\appendix

\section{Mappings for DIS}
\label{sec:DIS_mapping}

\subsection{Conventions}

The deep inelastic scattering process proceeds at LO via the parton reaction
\begin{equation}
  \label{eqn:Bornpartonlevel}
  \ell(l) + q(\bar{x} P) \to \ellp (\lp)+q\prime(v),
\end{equation}
where $P$ is the incoming nucleon momentum and, in general, the
final-state lepton and quark can be massive, $\lp^{2}= m_{\ellp}^{2}$
and $v^{2}= m_{v}^{2}$. In the following, we will always assume that
the nucleon is coming from the positive $z$-axis direction. The
corresponding Born phase space element is

\begin{align}
  \label{eqn:Bornphsp}
  \mathd{\bf{\Phi}}_B & =\mathd \bar{x}\frac{\mathd^3\lp}{2\lp^{0}(2\pi)^3}
                        \frac{\mathd^3v}{2v^0(2\pi)^3} (2\pi)^4\delta^{(4)}(l+\bar{x}P - \lp-v) \\
                      &= \mathd \bar{x} \frac{\mathd^3\lp}{4\lp^{0}v^{0}(2\pi)^2} \delta(\sqrt{\bar{x}S} - \lp^{0}-v^{0}) \\
                      &= \frac{1}{16\pi^{2}}  \mathd \bar{x} d \ydis d\phi_{\lp},
\end{align}
in terms of the $y\equiv \ydis$ and $\bar{x} \equiv \xbj$ variables
defined in eq.~\eqref{eqn:LIvariables}, 
$\phi_{\lp}$, the azimuthal angle of the outgoing lepton in a given reference
frame, and $S=(l+P)^{2}$ is the total energy of the lepton-nucleon collision.
At NLO, the real emission processes
\begin{align}
  \label{eqn:Bornpartonlevel}
  \ell(l) + q(x P) &\to \ellp (\lp)+q\prime(v) + g(k),\\
  \ell(l) + g(x P) &\to \ellp (\lp)+q\prime(v) + \bar{q}(k).
\end{align}
must be taken into account. The real emission phase space reads
\begin{equation}
  \label{eqn:Realphsp}
  \mathd{\bf{\Phi}}_R  =\mathd x\frac{\mathd^3\lp}{2\lp^{0}(2\pi)^3}
                        \frac{\mathd^3v}{2v^0(2\pi)^3}  \frac{\mathd^3k}{2k^0(2\pi)^3}(2\pi)^4\delta^{(4)}(l+\bar{x}P - \lp-v-k).
\end{equation}
The corresponding matrix elements develop singularities in the limit of soft
and/or collinear emission. In what follows we will show how to construct a
mapping from the real phase space ${\bf{\Phi}}_R$ to the underlying Born
variables (and its corresponding inverse mapping) for the two kinds of singular
regions, namely the initial and final state ones. We will consider mappings
compatible with the FKS~\cite{Frixione:1995ms}
subtraction method.

\subsection{Momentum mappings for DIS}

\subsubsection{DIS momentum mapping preserving the invariant mass of the born-like lepton-quark system}
\label{sec:simpisrmap}
We start by considering the mapping associated to the initial-state singular
region. We notice that the standard \POWHEG{} mapping introduced in ref.~\cite{Frixione:2007vw}
cannot be applied to the DIS kinematics as the longitudinal recoil of the
emitted parton is reabsorbed by both initial state partons. In this way, the
mapping preserves both the rapidity and the invariant mass of the Born system.
In this section, we will show a simple modification of that construction that
does not change the momentum of the incoming lepton.

Following the FKS formulation, we work in the rest frame of $l+ xP$, i.e.
the partonic CM frame of the real configuration, and we introduce the
standard FKS variables
\begin{equation}\label{eq:ISR-FKSvars}
  \quad \xi = \frac{2 k \cdot (l + p)}{(l + p)^2}, \quad y = \cos \theta_k, \quad \phi
\end{equation}
for the case of ISR radiation. The angles $\theta_k$ and $\phi$ are relative to
the positive beam axis direction. In this parametrisation, the momentum $k$ reads
\begin{equation}
  \label{eqn:gluonmomentum}
  k = \frac{\sqrt{xS}}{2} \xi (1,\sin{\theta}\sin{\phi},\sin{\theta}\cos{\phi},\cos{\theta}),
\end{equation}
and the corresponding one-particle phase space element is
\begin{equation}
  \label{eqn:phspgluon}
  \frac{\mathd^3 k}{2k^0(2\pi)^3}=\frac{x S}{(4\pi)^3}\xi\mathd\xi\mathd y\mathd\phi.
  \end{equation}
We introduce the momentum of the Born final state
\begin{equation}
    \label{eqn:app_ISRktot}
    k_{\rm{tot}}= l + x P-k =  \frac{\sqrt{x S}}{2}
    (2-\xi,-\xi\sin{\theta}\cos{\phi},-\xi\sin{\theta}\sin{\phi},-\xi\cos{\theta}).
\end{equation}
Following the standard \POWHEG{} construction we reabsorb the
longitudinal and transverse momentum by performing a sequence of boost transformations
\begin{equation}\label{eq:ISRmapBoost}
 \bar{k}_{\rm tot} =  B^{-1}_{\parallel}  B_{\perp} B_{\parallel}   k_{\rm tot},
\end{equation}
consisting in a longitudinal boost to the system of zero rapidity of
$k_{\rm tot}$, a transverse boost in order to absorb its transverse component
and, finally, the inverse of the longitudinal boost. By construction, the
invariant mass of the Born final-state system
$\bar{k}_{\rm tot}^{2}=k^{2}_{\rm tot}$ is preserved. We require now that
\begin{equation}
   \bar{k}_{\rm tot} = l + \bar{x} P,
\end{equation}
which, at variance with the standard \POWHEG{} construction, implies that we
cannot also preserve the rapidity of the Born system. From the conservation of
the invariant mass, we immediately obtain an expression of the underlying Born
momentum fraction $\bar{x}$ in terms of the radiation variables and $x$
\begin{equation}\label{eqn:app_ISRdirectmapp}
  (l+ x P-k)^2= (l+\bar{x} P)^2 \implies  x S (1-\xi) = \bar{x} S \implies \bar{x} = (1-\xi)x.
\end{equation}

The inverse mapping, consisting in reconstructing the kinematic of the real
emission from the underlying Born momenta and the radiation variables $\xi$, $y$
and $\phi$, can be easily obtained by inverting
eq.~\eqref{eqn:app_ISRdirectmapp}
\begin{equation}
    \label{eqn:app_ISRinversemapp}
    x = \frac{\bar{x}}{1-\xi},
\end{equation}
and eq.~\eqref{eq:ISRmapBoost}.

Requiring $x \leq 1$ we also obtain an upper bound for the energy fraction
of the emitted parton
\begin{equation}
    \label{eqn:app_ISRximax}
    \xi \leq 1-\bar{x} \equiv \xi_{\rm{max}}.
\end{equation}
The real emission phase space element reads
\begin{equation}
  d{\bf \Phi}_{R} = \mathd \bar{x} d\Phi_{B} \frac{\bar{x}S}{(4\pi)^{3}} \xi \mathd \xi \mathd y \mathd \phi \equiv  \mathd \bar{x} d\Phi_{B} J(\xi,y,\phi;\bar{x})  \mathd \xi \mathd y \mathd \phi
\end{equation}
from which we get the jacobian of the mapping
\begin{equation}
  J(\xi,y,\phi;\bar{x}) = \frac{\bar{x}S}{(4\pi)^{3}} \xi.
\end{equation}

\subsubsection{DIS momentum mapping preserving the lepton kinematics: the ISR case}
\label{sec:isr-dis-map}
A more interesting option is to set up a projection from a real phase space
configuration to an underlying Born one that preserves the momentum of the
scattered lepton. In this case, both $\ydis$ (or $Q^2$) and $\xbj$ remain
invariant, that is a natural choice for this process.

Consider the
system in the lepton-proton CM. Let us call $E$ the energy of the incoming
proton and lepton, $E'$ the energy of the outgoing lepton, and $\theta$ the
lepton scattering angle. Retaining the dependence on the mass of the outgoing
lepton, $m_{\ellp}$, we have
\begin{eqnarray}
  Q^2 &=& - (l - \lp)^2 = -m^2_{\ellp} + 2 l \cdot \lp = -m^2_{\ellp} + 2 E \Ep (1 - \bp \cos \theta),\\
  \nu &=& (l - \lp) \cdot P = 2 E^2 - E \Ep (1 + \bp \cos \theta),
\end{eqnarray}
where $\bp = |\vec{\lp}|/\Ep $. Then, it follows that both $\Ep$ and
$\theta$ must remain fixed to preserve $\xbj$ and $Q^2$. At the Born
level, the struck parton carries a fraction $\bar{x}$ of the incoming
proton momentum $P$. We consider the general case of producing a
massive final-state quark of momentum $v$ and mass $m_v$. By imposing
the on-shell condition on the scattered quark momentum, we get
$v=l - \lp + \bar{x} P = q + \bar{x}P$
\begin{equation}
  v^2 = - Q^2 + 2 \bar{x} \nu = m_v^2 \implies \bar{x} = \frac{Q^2+m_v^2}{2\nu}
\end{equation}
which reduces to $\xbj$ in the limit of a massless outgoing quark. If a gluon
with momentum $k$ is emitted, the momentum fraction $x$ associated to the real
configuration can be similarly computed starting from $v = q + xP - k$ and
imposing that $v$ is on the mass shell. We get
\begin{equation}\label{eq:xreal}
  x = \frac{Q^2 + m_v^2 + 2 k \cdot q}{2 (\nu - k \cdot P)}.
\end{equation}
In particular, we notice that if $k$ is collinear to $P$, say
$k = \xi p$, we have
\begin{equation}
  x = \frac{Q^2 + m_v^2 + 2 \xi \nu}{2 \nu} = \bar{x} + \xi,
\end{equation}
as expected.
Again, we work in the rest frame of $l+p$ with $p=xP$ and introduce standard FKS
variables as in eq.~\eqref{eq:ISR-FKSvars}. Then, the kinematics is given by
\begin{eqnarray}
  l &=& \frac{\sqrt{s}}{2} (1, 0_{\perp},  1), \\
  p &=& \frac{\sqrt{s}}{2} (1, 0_{\perp}, -1), \\ 
  k &=& \xi \frac{\sqrt{s}}{2} \left( 1, \sqrt{1 - y^2} s_\phi, \sqrt{1 -
        y^2} c_\phi, y \right)\;,
\end{eqnarray}
where $s = (l+p)^2$, $s_\phi\equiv \sin \phi$ and
$c_\phi\equiv \cos \phi$. We observe that the relevant initial state
collinear limit corresponding to the emitted gluon being collinear to
the incoming quark is approached for $y \to -1$. It is convenient to
write the momentum of the final-state lepton $\lp$ in terms of $Q^2$,
$\nu$ and $S=(l+P)^2$, which are preserved by the mapping, and of the
momentum fraction $x$. To this end, we start from the following
parametrisation
\begin{equation}
  \lp = \lp^0 (1, \bp \sin \tp, 0, \bp \cos \tp),
\end{equation}
where, without loss of generality, we set the azimuth of the
final-state lepton to zero. Then, the invariants $Q^2$ and $\nu$ can
be written as
\begin{eqnarray}
  Q^2 &=& -m_{\ellp}^2 + \sqrt{s} \lp^0 (1-\bp \cos \tp),\\
  2 \nu &=& S - \frac{\sqrt{s}}{x} \lp^0  (1+\bp \cos \tp).
\end{eqnarray}
Inverting the above system in terms of the variables $\lp^0$ and
$\bp \cos\tp$, we obtain
\begin{eqnarray}
  \lp^0 &=& \frac{Q^2+m_\ellp^2 +  (S-2\nu) x }{2\sqrt{s}} = \frac{Q^2+m_{\ellp}^{2}}{2\sqrt{s}}\left(1+\frac{x}{\chi} \right),\\
  \bp \cos \tp &=& 1 - 2\frac{Q^2+m_\ellp^2}{Q^2+m_\ellp^2 + (S-2\nu)x} = \frac{x-\chi}{x+\chi}.
\end{eqnarray}
where we have introduced the quantity
\begin{equation}
  \chi \equiv \frac{Q^2+m_\ellp^2}{S-2\nu} = x \frac{1 - \bp \cos\tp }{1 + \bp \cos \tp }.
\end{equation}
In summary, we have
\begin{equation}
  \lp = \frac{Q^2+m_\ellp^2}{2 \chi \sqrt{s}} \left[ x + \chi, 2 \sqrt{\chi x -  \frac{ \chi^2 m_\ellp^2 s}{(Q^2+m_\ellp^2)^2 }}, 0,
    x- \chi \right].
\end{equation}
The momentum of the final-state quark is then fixed by momentum
conservation to be $v=q+p-k$. In this way, we have constructed a
mapping from a given Born phase space point and radiation variables
$\xi,y,\phi$ to a real one under the assumption that
eq.~\eqref{eq:xreal} admits at least one solution. Indeed, $x$ must
satisfy the equation
\begin{eqnarray}\label{eq:xreal_master}
0 &=& \left[\St \xi (1+y) - 4\nu (1-\xi) \right]x + \left[ 2s_\phi \xi \sqrt{\left(\St Q^2-2\nu m_\ellp^2\right)(1-y^2)}  \right]\sqrt{x}  \nonumber \\ &&+ 4\nu \bar{x}- (Q^2+m_\ellp^2)\xi(1+y)
  \equiv \ax x + \bx \sqrt{x} + \cx .
\end{eqnarray}
where $\St = S-2\nu$. Notice that the dependence on the mass of the
final-state quark is implicitly contained in the Born momentum
fraction $\bar{x}$. In the soft limit $\xi = 0$, the equation becomes
\begin{equation}
  4\nu (-  x + \bar{x} ) = 0 \Rightarrow x = \bar{x}.
\end{equation}
as it should. We look after solutions in the physical range
$[\bar{x},1]$ of eq.~\eqref{eq:xreal_master}, which is quadratic in
$\sqrt{x}$. At fixed underlying Born configuration, this leads to non
trivial boundaries in the radiation phase space
$[0,1]_\xi\times[-1,1]_y\times[0,2\pi]_\phi$. In order to have real
solution, the discriminant of eq.~\eqref{eq:xreal_master} must be
positive:
\begin{eqnarray}\label{eq:discriminant}
  \Delta_x &=& 4\left[(\St Q^2-2m_\ellp^2\nu)s_\phi^2(1-y)+(Q^2+m_\ellp^2)(4\nu+\St(1+y))\right](1+y) \xi^2 \nonumber \\
           &&-16\nu \left[ 4\nu\bar{x} + (Q^2+m_\ellp^2+\St\bar{x})(1+y)\right]\xi + 64\nu^2\bar{x}  \nonumber \\
           &\equiv& \axi\xi^2 + \bxi\xi + \cxi > 0 
\end{eqnarray}
We employ the above condition to derive constraints on the $\xi$
variable as function of the underlying Born and the other two
radiation variables $y$ and $\phi$. Eq.~\eqref{eq:discriminant} is
quadratic in $\xi$ with a clearly positive coefficient of the $\xi^2$
term as the quantity $(\St Q^2-2m_\ellp^2\nu) \ge 0$ in the $2\to2$
Born kinematics. In order to study if there are real solutions, we
evaluate its discriminant
\begin{eqnarray}\label{eq:discriminant-xi}
  \Delta_\xi &=& (16\nu )^2\left[ ( (Q^2+m_\ellp^2 - \St\bar{x})(1+y) - 4\nu\bar{x})^2  - 4s_\phi^2 \bar{x}(\St Q^2-2m_\ellp^2\nu)(1-y^2)   \right] \nonumber \\
             &\ge&  (16\nu )^2\left[ ( (Q^2+m_\ellp^2 - \St\bar{x})(1+y) - 4\nu\bar{x})^2  - 4 \bar{x} (\St Q^2-2m_\ellp^2\nu)(1-y^2)   \right],
\end{eqnarray}
being the coefficient of $s_\phi^2$ negative. The quantity in the square
bracket is a quadratic polynomial in $y$ that is equal to $16 \nu^2 x_b^2>0$ at
$y=-1$ and, as its discriminant is negative
\begin{equation}
\Delta_y = -64 S \bar{x}^2 m_v^2(\St Q^2-2m_\ellp^2\nu) < 0,
\end{equation}
it is positive definite. Therefore, we conclude that $\Delta_\xi>0$,
there are always two real solutions $\xi_1<\xi_2$ and $\Delta$ is
positive for $\xi\le\xi_1$ or $\xi\ge\xi_2$ in the allowed range
$[0,1]_\xi$. Furthermore:
\begin{enumerate}
\item the coefficients of the quadratic polynomial in $\xi$ in
  eq.~\eqref{eq:discriminant} have definite signs and it follows that
  $\xi_1>0$  ( $\axi>0$, $\bxi<0$ and $\cxi>0$ );
\item the upper limit for $\xi=1$ corresponds to the condition
  $\axi+\bxi+\cxi = 0$. Solving it with respect to $y$, we
  obtain two solutions
\begin{equation}
  y_1=-1, \quad \quad
  y_2=1 + 2 \frac{ \St (m_v^2-m_\ellp^2)}{ \St (Q^2 + m_\ellp^2) -
    (\St Q^2 - 2m_\ellp^2\nu)s_\phi^2}.
\end{equation}
\item We consider first the solution $y_1$. We observe that
  $\xi_1 = 1$ at $y=y_1$ independently of the Born kinematics and
  $\phi$. Being $\xi_2>\xi_1$, the second solution leads to a non
  physical value for $\xi$.  It turns out that $\xi_1$ (and $\xi_2$)
  is a decreasing function of $y$ since the derivative of $\xi_1$ with
  respect to $y$ never vanishes and $\xi_1(y_1) = 1$ and $\xi_1\to0$
  in the limit $y\to +\infty$. In fact, the discriminant $\Delta_{dy}$
  associated to the equation $d\xi_1/ dy = 0$ is negative
  \begin{eqnarray}
    \Delta_{dy} &\propto&  2\nu(Q^2+m_\ellp^2 - (\St+2\nu)\bar{x}) + (\St Q^2 - 2m_\ellp^2\nu)s_\phi^2  \\
    &<&  2\nu(Q^2+m_\ellp^2 - (\St+2\nu)\bar{x}) + (\St Q^2 - 2m_\ellp^2\nu) = - m_v^2 ( \St + 2\nu) < 0\,, \nonumber
  \end{eqnarray}
  where we have factored out a positive quantity.  We conclude that the solution
  \mbox{$0<\xi_1<1$} is always allowed.

\item Then, it clearly follows that $\xi_2=1$ at $y=y_2$. We notice
  that, in the case $m_v \ge m_\ellp$, $y_2\ge 1$. Then, $\xi_2>1$ in
  the physical region and must be discarded. For $m_v<m_\ellp$,
  $y_2 < 1$ and the solution $\xi_2$ is allowed for values of $y$ in
  the range $[y_2,1]$, independently of the values of $\phi$.

\end{enumerate}

We turn now to discuss the physical solutions of eq.~\eqref{eq:xreal_master}.
Formally, eq.~\eqref{eq:xreal_master} admits the two solutions
\begin{equation}
  \sqrt{x_{1,2}} = \frac{-\bx\pm\sqrt{\bx^2-4\ax\cx}}{2\ax}.
\end{equation}
A physical solution corresponds to $\bar{x}<x<1$. We need to consider two
limiting cases: one associated to the solution crossing $x=1$, and the other to
the limit in which the coefficient $\ax$ is vanishing, where one of the two
solutions develops a singularity.

Concerning the latter, we observe that $\ax$ is a linear and increasing function
of $\xi$; solving the equation $\axi=0$ as function of $\xi$ we find
\begin{equation}
  \xi_\ax = \frac{4\nu}{4\nu+\St(1+y)}.
\end{equation}
Since the discriminant $\Delta_{x}$ in
  eq.~\eqref{eq:discriminant} does not vanishes when $\ax=0$, it follows that
$\xi_\ax$ is always in the physical region. Moreover, $\xi_\ax$ is well defined
for every $y$ and, being a continuous function of $y$, it must be
$\xi_\ax< \xi_1$. The solution $x$ crosses $x=1$ when $\ax+\bx+\cx=0$. Solving
this equation with respect to $\xi$, we get
\begin{equation}
  \xis =  \frac{4\nu(1-\bar{x})}{4\nu + (\St-Q^2-m_\ellp^2)(1+y) + 2s_\phi\sqrt{(\St Q^2-2m_\ellp^2\nu)(1-y^2)}},
\end{equation}
and $\xis<\xi_1$ following the same reasoning as for $\xi_\ax$. 
In order to discuss the solutions, we separate the regions $\xi<\xi_1$
and $\xi_2<\xi<1$. In the former, the condition $\cx>0$ holds. It
follows that for $\xi<\xi_\ax$, $\ax<0$ and the quantity
$\sqrt{\bx^2-4\ax\cx} >|\bx|$. Since $x_2=\bar{x}$ is the valid
solution in the soft limit $\xi \to 0$, for continuity it must be a
valid solution in some subset of the interval $0<\xi<\xi_\ax$,
regardless the sign of $\bx$ which, in turn, only depends on the sign
of $s_\phi$.  Then it follows that
\begin{enumerate}
  
\item for $\xi_\ax>\xis$, there is only one solution corresponding to
  $\sqrt{x_2}$ for $0<\xi<\xis$, as $\sqrt{x_2}$ reaches $1$ at
  $\xi=\xis$. In fact, the other solution $\sqrt{x_1}$ is negative at
  $\xi=0$ and, since it cannot vanish, stays negative in the
  considered range.

\item for $\xi_\ax<\xis$, the situation is more involved. The solution
  $x_2$ must be continuous at $\xi=\xi_\ax$, which implies that
  $\bx<0$, and must be valid at least up to $\xis$ where it might
  become $1$. The other solution $x_1 \to +\infty$ as
  $\xi \to \xi_\ax^-$ and stays greater than $1$ at least up to
  $\xis$. At $\xis$, one of the two solutions $x_1$ and $x_2$ becomes
  $1$. If it is not $x_2$, then $x_2$ would be a valid solution up to
  edge $\xi_1$ where $x_2= -\frac{\bx}{2\ax}|_{\xi=\xi_1}$. This
  requires that the quantity $-\frac{\bx}{2\ax}|_{\xi=\xi_1} \le
  1$. In this case, then
  \begin{itemize}
  \item the solution $x_2$ is valid for $0<\xi<\xi_1$; 
  \item the solution $x_1$ is valid for $\xis < \xi < \xi_1$. 
  \end{itemize}
  and there are two possible solutions in the region $\xis < \xi < \xi_1$.  On the other
  hand, if $-\frac{\bx}{2\ax}|_{\xi=\xi_1} \ge 1$, it must be the
  solution $x_2$ to cross 1 at $\xis$. Then, it follows that there is
  only one acceptable solution, $x_2$, for $0<\xi<\xis$.  
\end{enumerate}

It follows from the above discussion that, in particular, in the
region specified by the conditions $\xi_\ax<\xis$,
$-\frac{\bx}{2\ax}|_{\xi=\xi_1} \le 1$ and $\xis < \xi < \xi_1$, both
solutions $x_{1,2}$ are allowed and lead to two different physical
real configurations. In this region, therefore, the mapping is not
bijective and there are two possible branches, which are present also
in the limit of massless lepton and quark, as observed in
Ref~\cite{Banfi:2023mhz}. Finally, in the region $\xi_2<\xi<1$, it
turns out that $x_{1,2}$ are both outside the physical region
$[\bar{x},1]$, and thus there are no physical solutions.

\begin{figure}
  \centering
  \includegraphics[width=0.45\textwidth]{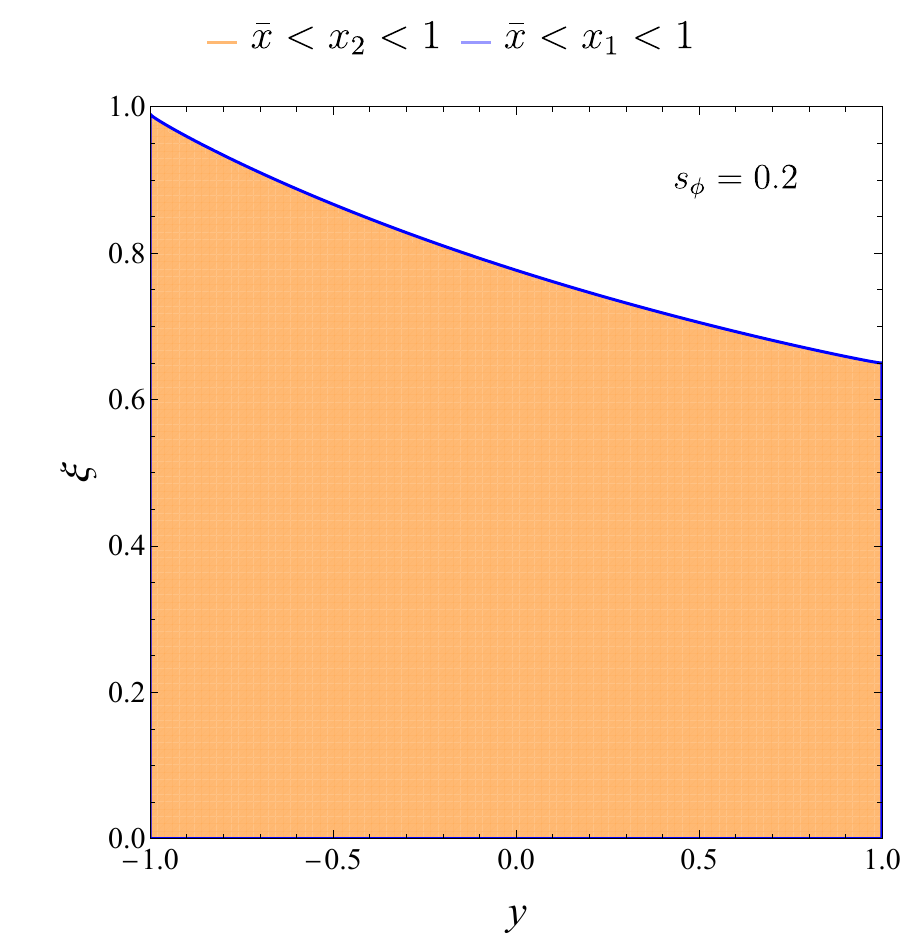}\hfil
  \includegraphics[width=0.45\textwidth]{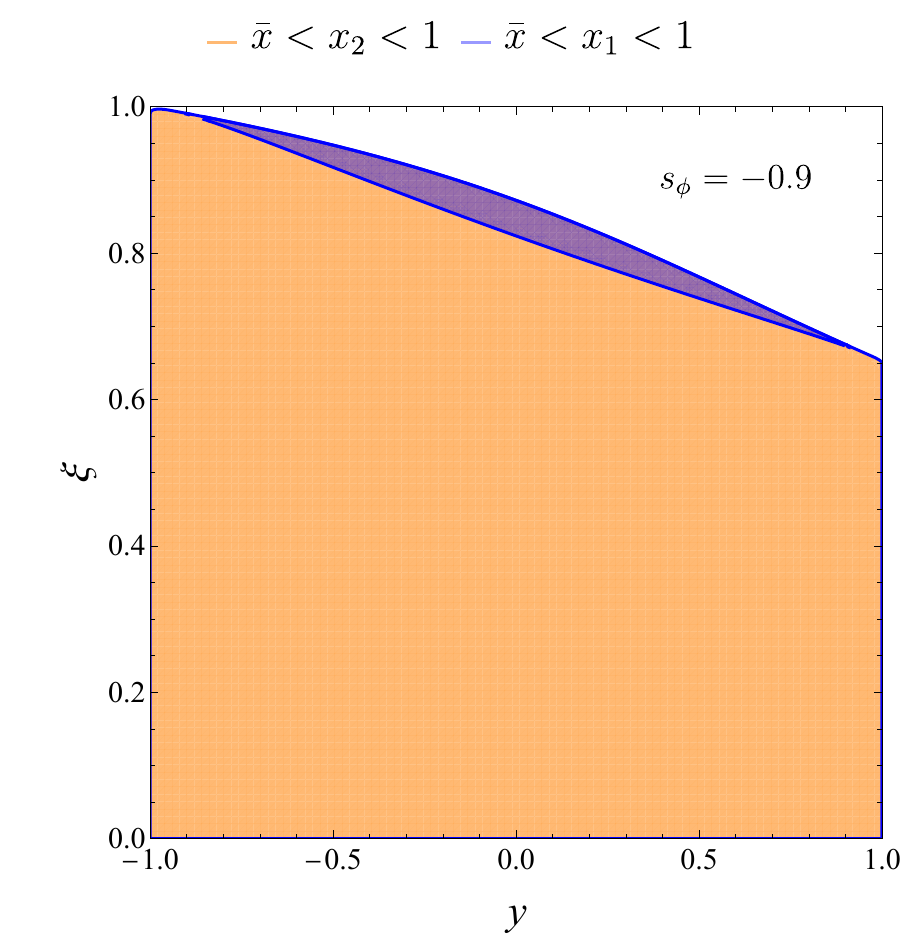}\hfil
  \caption{\label{fig:ISR_physical_region} Physical regions in the $(y,\xi)$
    radiation plane at fixed underlying Born configuration, given by
    $S=1876\,{\rm GeV}^{2}, \bar{x}=0.01, Q^{2}=10\,{\rm GeV}^{2}$, for
    $s_{\phi}=0.2$ (left panel) and $s_{\phi}=-0.9$ (right panel).}
\end{figure}

For illustrative purposes, we show the physical region in the
$(y,\xi)$ radiation plane in figure~\ref{fig:ISR_physical_region}, at a
fixed underlying Born configuration given by $S=1876\,{\rm GeV}^{2},
\bar{x}=0.01, Q^{2}=10\,{\rm GeV}^{2}$. The doubly-covered region
occurs for negative and large $s_{\phi}$ values.

\subsubsection{Jacobian for ISR mapping}
We write the real phase space as
\begin{equation}
  \mathd {\bf \Phi}_{R} = \mathd x \frac{\mathd^4 \lp}{(2\pi)^3} \delta \left( {\lp^2-m_\ellp^2} \right)
  \frac{\mathd^4 k}{(2\pi)^3}  \delta (k^2) 2 \pi \delta (v^2-m_v^2), \quad v = x P + l - \lp - k.
\end{equation}
Adopting FKS variables for $k$, we have
\begin{equation}
\frac{\mathd^4 k}{(2\pi)^3} \delta (k^2) = \frac{s}{(4 \pi)^3} \xi \mathd \xi \mathd y
\mathd \phi.
\end{equation}
We perform the integration over $x$ by solving the
$\delta (v^2-m_v^2)$. To this end, taking into account eq.~\eqref{eq:xreal}
and eq.~\eqref{eq:xreal_master}, we write
\begin{equation}
  \delta(v^2 - m_v^2) = \delta\left( \frac{1}{2} (\ax x + \bx \sqrt{x} + \cx)\right) = 2 \delta\left( \ax x + \bx \sqrt{x} + \cx\right). 
\end{equation}
The Jacobian from the $x$ integration is given by
\begin{eqnarray}
  J_x =  \int \mathd x 2 \delta\left(\ax x + \bx \sqrt{x} + \cx \right)
  &=& \int 4 y \mathd y \frac{1}{|\ax|} \delta\left( (y-\sqrt{x_1})(y-\sqrt{x_2}) \right) \nonumber \\
  &=& \int 4 y \mathd y \frac{1}{|\ax|} \delta\left( (y-\sqrt{x_1})(y-\sqrt{x_2}) \right) \nonumber  \\
  &=& \int 4 y \mathd y \frac{1}{|\ax|} \frac{1}{|\sqrt{x_1}-\sqrt{x_2}|} \left[\delta(y-\sqrt{x_1}) + \delta(y-\sqrt{x_2}) \right] \nonumber  \\
  &=& \frac {4 \sqrt{x_{1,2}}} { \sqrt{\Delta_x} }
\end{eqnarray}
where we have performed the change of variable $y=\sqrt{x}$, and the numerator $\sqrt{x_{1,2}}$ must be chosen according to the physical solutions.

Finally, we get
\begin{eqnarray}
  \mathd {\bf \Phi}_{R} &=& J_x \frac{s}{(4 \pi)^3} \xi \mathd \xi \mathd y \mathd \phi \; 2\nu \times \frac{1}{2\nu} \frac{\mathd^4 \lp}{(2\pi)^2} \delta \left( {\lp^2-m_\ellp^2} \right) \nonumber \\
              &=& 2\nu J_x \frac{s}{(4 \pi)^3} \xi \mathd \xi \mathd y \mathd \phi \times \mathd {\bf \bar{\Phi}}_{B} \equiv J(\xi,y,\phi) \mathd \xi \mathd y \mathd \phi \times \mathd {\bf \bar{\Phi}}_{B}
\end{eqnarray}
where $\mathd {\bf \bar{\Phi}}_{B}$ is the phase space element of the underlying
Born. The Jacobian associated to the radiation variable is then 
\begin{equation}
  J(\xi,y,\phi) =     \frac{\nu s}{(2\pi)^3} \xi  \sqrt{ \frac {x_{1,2}}{ \Delta_x }}.
\end{equation}

\subsubsection{Generation of ISR radiation}
\label{sec:genISRrad}
In the following, we highlight the main complications occurring at the
generation stage, when the events are generated according to the \POWHEG{}
method. First, we recall that the hard scale $K_{T}(\mathbf{\Phi})_{R}$ for
initial-state radiation is by default chosen to be the transverse momentum of the radiated
parton $k_{T}^{2}$. For the original ISR mapping implemented in the \POWHEGBOX{},
$k_{T}^{2}$ assumes a relatively simple form in terms of the radiation variables
$\xi,y,\phi$, i.e.
\begin{equation}\label{eq:kt2}
k_{T}^{2}= \frac{s}{4}\xi^{2}(1-y^{2}) = \frac{\bar{s}}{4(1-\xi)}\xi^{2}(1-y^{2})
\end{equation}
where $s$ and $\bar{s}$ are the partonic centre-of-mass energies in the real and
the underlying Born configuration, respectively. While eq.~\eqref{eq:kt2}
holds also for the case of the
DIS mapping that preserves the invariant mass of the Born system (discussed in
section~\ref{sec:simpisrmap}), the situation is more involved for the new DIS
mapping preserving the lepton momenta derived in the previous section. Now the
relation between $s$ and $\bar{s}$ involves the real momentum fraction
$x = x(\xi,y,\phi)$, which is a complicated function of the radiation variables:
\begin{equation}\label{eq:kt2new}
k_{T}^{2}= \frac{s}{4}\xi^{2}(1-y^{2}) = \frac{ x(\xi,y,\phi) \bar{s}}{4 \bar{x}} \xi^{2}(1-y^{2}).
\end{equation}
The construction for the generation of initial-state radiation in \POWHEG{} is
based on the appendix of ref.~\cite{Nason:2006hfa}, which in turn relies on the
particular expression of $k_{T}^{2}$ in the right hand side of
eq.~\eqref{eq:kt2}. Therefore, the standard implementation provided by the
\POWHEGBOX{} is inconsistent with the new DIS mapping. On the other hand, one
can choose the \POWHEG{} hard scale in a different way as long as it reduces to
the transverse momentum of the radiated parton in the relevant soft and/or
collinear limits. We exploit this freedom and define the hard scale as (in our
convention the collinear singularity is approached only for $y\to -1$):
\begin{equation}\label{eq:kt2isr}
K_{T}^{2}(\mathbf{\Phi}_{R})= \frac{\bar{s}}{2}\xi^{2}\frac{1+y}{1+\xi y}
\end{equation}
and we use as upper bound function for the veto method
\begin{equation}\label{eq:uboundisr}
U_b= \frac{N\alpha_s(k_{T})}{\xi (1+y)}.
\end{equation}
The integration of the upper bound (including the lowest-order running for
$\alpha_s$) times the factor $\theta(t-k_T)$ is straightforward and the veto
technique is used to generate the hardest radiation according to the \POWHEG{}
formula in eq.~\eqref{eq:radmaster}. We point out that we have chosen to use the hard scale
in eq.~\eqref{eq:kt2isr} and the upper bound function in
eq.~\eqref{eq:uboundisr} also for the mapping preserving the invariant mass of
the Born system and the momentum of the initial lepton. Differences among
various options for the hard scale are known to have impact on the resummed
results only at higher orders.

A second problem is associated with the region where the DIS mapping
is not invertible, i.e. in the region where there are two physical
real configurations associated with the same underlying Born
$\bar{\mathbf{\Phi}}_{B}$ and set of radiation variables
$\xi,y,\phi$. We notice that the problematic region does not encompass
any of the singular limits. Therefore, it can be removed from the real
contributions that appear in the \POWHEG{} formula and can be treated
separately exploiting the standard remnant mechanism provided by the
\POWHEGBOX{}.

\subsubsection{DIS momentum mapping preserving the lepton kinematics: the FSR case}
\label{sec:fsr-dis-map}
We consider now the case of a final-state singular region and we focus on a
mapping that preserves the lepton kinematics. We observe that this mapping is
actually the same as the one described in Sec.~\ref{sec:isr-dis-map} for the
initial-state case. Nonetheless the meaning of the FKS radiation variables is
different in the two cases and thus a dedicated construction is needed. In this
section, we must assume that the final-state emitter quark is massless, since in
\POWHEG{} no mapping is associated to massive quarks. We instead retain the
possibility of having a massive outgoing lepton.

The construction proceeds as follows. We work in the partonic CM frame and, as
first step, we integrate out the quark momentum by exploiting the 3-momentum
conservation. We adopt a parametrisation of the momenta in terms of spherical
coordinates
\begin{eqnarray}
  \mathd \mathbf{\Phi}_{R} & = & \mathd x \frac{\mathd^3 \lp}{(2 \pi)^3 2 \lp^0} \frac{\mathd^3 v}{(2 \pi)^3 2 v^0}
               \frac{\mathd^3 k}{(2 \pi)^3 2 k^0} (2 \pi)^4 \delta^4 (l + x P - \lp - v - k) \notag \\
         & = & \mathd x \frac{1}{8 (2 \pi)^5} \frac{\mathd^3 \ulp \mathd^3 k}{\lp^0 v^0 k^0} \delta
               \left( \sqrt{s } - \lp^0 - v^0 - k^0 \right) \notag \\
         & = & \mathd x \frac{1}{8 (2 \pi)^5} \frac{\ulp^2 \mathd \ulp  \mathd \cos \theta_\lp \mathd
               \phi_\lp}{\lp^0 v^0} k^0 \mathd k^0 \mathd c_{\psi} \mathd \phi \notag \\  && \times \delta \left( \sqrt{s } - \lp^0 -
               \sqrt{\ulp^2 + (k^0)^2 - 2 \ulp  k^0 c_{\psi}} - k^0 \right),
\end{eqnarray}
where $\ulp$ denotes the modulus of the tri-momentum of $\lp$ and
$c_{\psi}\equiv \cos \psi$. We use, as reference axes for the angles
$\theta_{\lp}$ and $\psi$, the direction of the incoming lepton and that
opposite to the outgoing one, respectively. We notice that we cannot use
directly the direction of the outgoing quark as it has been integrated out.

We parametrise the lepton variables in terms of the DIS invariants
\begin{eqnarray}
  Q^2 & = & \sqrt{s} (\lp^0 - \ulp \cos \theta_{\lp}) - m_{\ellp}^2\,,\\
  \ydis & = & 1 - \frac{\lp^0 +  \ulp \cos \theta_{\lp}}{\sqrt{s}}\,.
\end{eqnarray}
This is convenient in order to make transparent the connection to a Born
configuration, obtained by absorbing the recoil of the radiation. In particular,
the CM energy is not preserved and it changes from $s$ to $s_b = \lambda s$,
going from a real to a Born configuration, and the partonic Born CM frame does
not coincide with the real CM one, as it is the case in the standard final-state
FKS mapping. The advantage of using DIS invariants rests on the fact that those
variables are frame independent.

\begin{figure}
  \centering
  \includegraphics[width=0.75\textwidth]{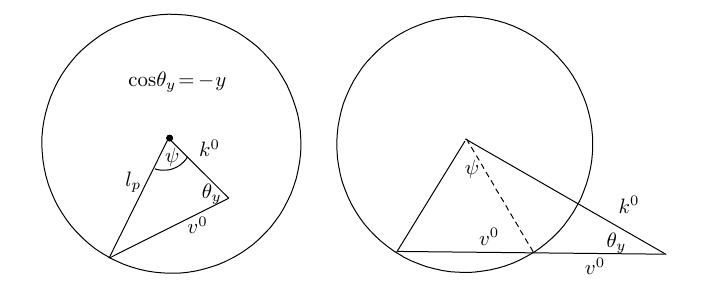}
  \caption{\label{fig:ycpsi-inv} Geometric illustration of the relation between
    the angle $\psi$ and the standard FKS one $\pi-\theta_{y}$ for the two cases
    $\ulp > k^0$ (left) and $\ulp < k^0$ (right).}
\end{figure}

Computing the jacobian associated to the above 2-dimensional change of
variables, we get
\begin{eqnarray}
  \tmop{jacobian}^{- 1} & = & \left| \frac{\partial (Q^2,
  \ydis)}{\partial (\ulp, \cos \theta_\lp)} \left| = \left|
  \begin{array}{ll}
    \sqrt{s} \left( \frac{\ulp}{\lp^0} - \cos \theta_\lp \right) & -
    \frac{1}{\sqrt{s}}  \left( \frac{\ulp}{\lp^0} + \cos \theta_\lp \right)\\
    - \sqrt{s} \ulp & - \frac{1}{\sqrt{s}} \ulp
  \end{array} \right|  = 2 \frac{\ulp^2}{\lp^0}\,. \right. \right.
\end{eqnarray}
Then, we get
\begin{equation}
  \mathd \mathbf{\Phi}_{R} =  \mathd x \frac{1}{16 (2 \pi)^5} \frac{\mathd Q^2 \mathd
    \ydis \mathd \phi_\lp}{v^0} k^0 \mathd k^0 \mathd c_{\psi} \mathd
     \phi \delta \left( \sqrt{s } - \lp^0 - v^0 - k^0 \right) .
\end{equation}
In order to make contact with the FKS parametrisation, we want to pass from
$c_{\psi}$ to the FKS variable $y$, that is the cosine of the angle between $l'$
and $v$. Using Carnot's formula, we have
\begin{equation}
  y = \frac{\ulp^2 - v_0^2 - (k^{0})^2}{2 v^0 k^0} = \frac{2 c_{\psi} \ulp - \sqrt{s} \xi}{\sqrt{4 \ulp^2 - 4 c_{\psi}  \ulp \sqrt{s} \xi + s \xi^2}},
\end{equation}
with $v^0 = \sqrt{\ulp^2 + (k^{0})^2 - 2 c_{\psi} \ulp k^0}$.
The derivative with respect to $c_{\psi}$
\begin{equation}
  \frac{\mathd y}{\mathd c_{\psi}}  =  \frac{{4 \ulp}^2}{\left( 4 \ulp^2 - 4 c_{\psi}  \ulp \sqrt{s} \xi + s
  \xi^2 \right)^{3 / 2}} \left( 2 \ulp - c_{\psi} \sqrt{s} \xi \right),
\end{equation}
is positive for $\ulp > k^0$. In this case, $y$ is a monotonically increasing
function of $c_{\psi}$ and admits an unique inverse. The inverse can be obtained
by taking the square and solving the resulting quadratic equation for $c_{\psi}$
\begin{eqnarray}
  y^2 & = & \frac{\left( 2 c_{\psi} \ulp - \sqrt{s} \xi \right)^2}{4 \ulp^2 - 4
  c_{\psi} \ulp \sqrt{s} \xi + s \xi^2}, \nonumber\\
  \implies c_{\psi} & = & \frac{\pm y \sqrt{4 \ulp^2 - s \xi^2 (1 - y^2)} + \sqrt{s} \xi
  (1 - y^2)}{2 \ulp}.
\label{eq:cpsiy}
\end{eqnarray}
For $\ulp > k^0$, we must take the upper sign, that guarantees that $c_{\psi}$
is a monotonically increasing function of $y$ mapping the $[- 1, 1]$ interval
biunivocally into itself. For $\ulp < k^0$ both signs are acceptable. The
possible cases are depicted in figure~\ref{fig:ycpsi-inv}, where we exploit the
geometrical relation among the angles $c_{\psi}$ and
$y = \cos (\pi - \theta_y) = -\cos (\theta_y)$ and the lengths $v^0$, $k^0$ and
$\ulp$ forming a triangle. The left panel corresponds to the case $\ulp > k^0$.
It is clear that there is a one-to-one correspondence between $c_{\psi}$ and $y$.

On the other hand, for the case $k^0 > \ulp$ shown in the right panel, the range
of $\theta_y$ corresponding to $c_{\psi} \in [-1,1]$ does not fully cover the
interval $[0,\pi]$ and in particular $\theta_y$ goes from 0 up to a maximum value
strictly lower than $\pi$ such that the collinear region is excluded. Furthermore,
because the transverse momentum of the two final state partons has to compensate the
one of the lepton, radiation has to be necessarily in the anticollinear region,
at $y<0$.
This region can be covered by the remnant mechanism, since it is not singular.
Correspondingly, for $\ulp < k^0$,
we must have $y < 0$ and
\begin{equation}
  v^0 = \frac{1}{2} \left( \pm \sqrt{4 \ulp^2 - s \xi^2 (1 - y^2)} - \sqrt{s}
    \xi y \right),
\end{equation}
while for $\ulp > k^0$, we have
\begin{equation}
v^0 = \frac{1}{2} \left( \sqrt{4 \ulp^2 - s \xi^2 (1 - y^2)} - \sqrt{s} \xi
  y \right) .
\end{equation}
The choice of the sign guarantees that for $y = - 1$
\begin{equation}
  v^0 = \frac{1}{2} \left( 2 \ulp + \sqrt{s} \xi \right) = \ulp + k^0\,,
\end{equation}
and for $y = 1$
\begin{equation}
  v^0 = \frac{1}{2} \left( 2 \ulp - \sqrt{s} \xi \right) = \ulp - k^0\,.
\end{equation}
Using the relation \ref{eq:cpsiy} we can compute the jacobian factor associated
with the change of variable $c_{\psi} \to y$, and we get
\begin{equation}
  J_y = \left| \frac{4 \sqrt{s} \xi y v^0 - (4 \ulp^2 - s \xi^2)}{2 \ulp
  \sqrt{4 \ulp^2 - s \xi^2 (1 - y^2)}} \right| = \frac{(v^{0})^{2}}{\ulp|v^{0}+k^{0}y|}\,,
\end{equation}
where we have conveniently identified a factor $v_0$ to simplify the notation. Then, we have
\begin{equation}
  \mathd \mathbf{\Phi}_{R} =  \mathd x \frac{1}{16 (2 \pi)^5} \frac{\mathd Q^2 \mathd
    \ydis \mathd \phi_\lp}{v^0} k^0 \mathd k^0 \mathd y \mathd
     \phi \, J_y \delta \left( \sqrt{s } - \lp^0 - v^0 - k^0 \right) .
\end{equation}
The last step in the construction of the full jacobian factor is to perform
the change of variable $x=s/S$ and integrate in $\mathd s$ exploiting the $\delta$ function.
As a result, the $\delta$ function enforces the relation
\begin{equation}
  s =  \frac{2 Q^2 - (Q^2 + m_{\ellp}^2) \xi (1 - y)}{\ydis (2 - \xi (1 -
  y)) - \xi (1 - \xi) (1 - y)}
\end{equation}
and after using the properties of the Dirac $\delta$ and a bit of algebra we have the last
jacobian factor
\begin{equation}
J_{\delta}  =  \left| \frac{4 \sqrt{s}  \left( 2 s - 2 \sqrt{s} l^0 - s
  \xi (1 - y) \right) /s}{(4 + 2 (\xi^2 - 2 \xi) (1 - y) - 2 (1 - \ydis)
  (2 - \xi (1 - y)) )} \right|
= \frac{4|v^{0}+k^{0}y|}{2 \ydis - \xi (1 + \ydis - \xi) (1 - y)}
\end{equation}
so that the final expression reads
\begin{eqnarray}
  \mathd \mathbf{\Phi}_R &=& \frac{1}{S} \frac{1}{16 (2 \pi)^5} \frac{\mathd Q^2 \mathd
    \ydis \mathd \phi_{\lp}}{v^0} k^0 \mathd k^0 \mathd y \mathd \phi
  J_y J_{\delta} \\ &=& \frac{1}{S} \frac{1}{4 (2 \pi)^5} \frac{k^0v^{0} \mathd Q^2 \mathd
  \ydis \mathd \phi_{\lp} \mathd k^0 \mathd y \mathd \phi}{\ulp(2 \ydis - \xi (1 + \ydis - \xi) (1 - y))} \\ &=& \frac{\ydis}{2 \ydis - \xi (1 + \ydis - \xi) (1 - y)}\frac{2 v^{0}}{\ulp} \frac{\mathd^{3}k}{(2\pi)^{3}2k^{0}} \times \mathd \mathbf{\bar{\Phi}}_{B}
\end{eqnarray}

\subsubsection{Generation of FSR radiation}
\label{sec:genFSRrad}
Being the map among real and born configuration essentially the same for FSR and
ISR, similar considerations apply for the definition of the hard scale for
FSR. The original choice done in the \POWHEGBOX{} framework are slightly
modified to consistently address the case of DIS. We use as evolution variable
\begin{equation}\label{eq:kt2fsr}
K_{T}^{2}(\mathbf{\Phi}_{R})= \frac{\bar{s}}{2}\xi^{2}(1-y)\,,
\end{equation}
and as upper bound function
\begin{equation}\label{eq:uboundfsr}
  U_b= \frac{N\alpha_s(k_{T})}{\xi (1-y)}\,.
\end{equation}
Note that for the case of final state radiation in our mapping the collinear
region is approached in the limit $y\to 1$, and the upper bound function is the
same as for the case of initial state radiation.

\section{Further NLO validation plots}
\label{sec:NLO_val_plot}

For convenience we show here our validation plots for other relevant cases of
lepton-hadron DIS processes. The setup, cuts and scale setting are the same as
reported in section~\ref{sec:nloval}.
\begin{figure}[htb]
  \includegraphics[width=0.45\textwidth,page=13]{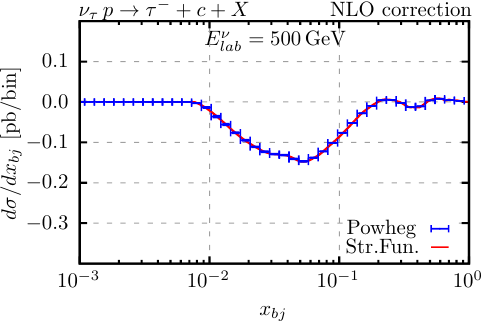}
  \includegraphics[width=0.45\textwidth,page=13]{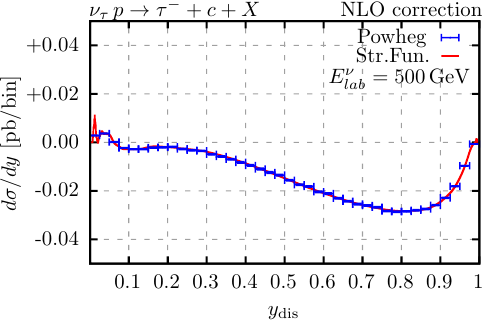}
  \includegraphics[width=0.45\textwidth,page=9]{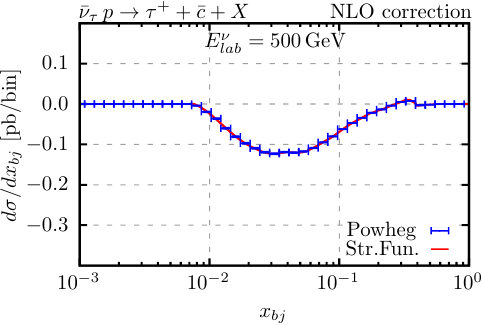}
  \hspace{1.25cm}
  \includegraphics[width=0.45\textwidth,page=9]{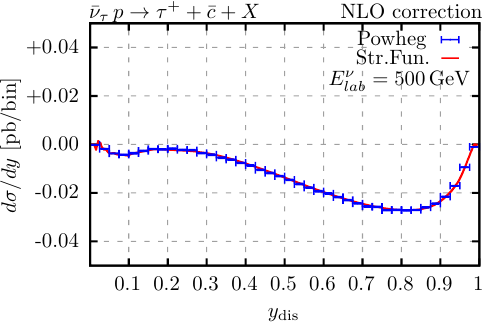}
  \caption{Same as figure~\ref{fig:CCnxy} for charged current $\nu_\tau$ ($\bar{\nu}_\tau$) DIS
    with $m_\tau = 1.777\,$GeV and charm quark production setting $m_c = 1.5\,$GeV.
  }
\label{fig:nCCtauc}
\end{figure}

\begin{figure}
  \includegraphics[width=0.45\textwidth,page=1]{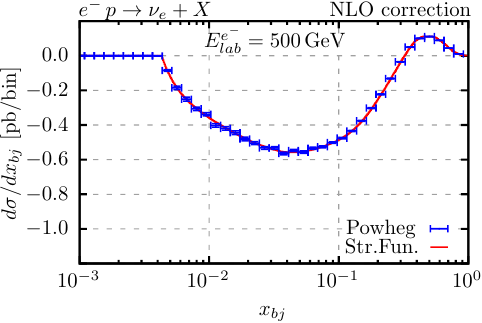}
  \includegraphics[width=0.45\textwidth,page=1]{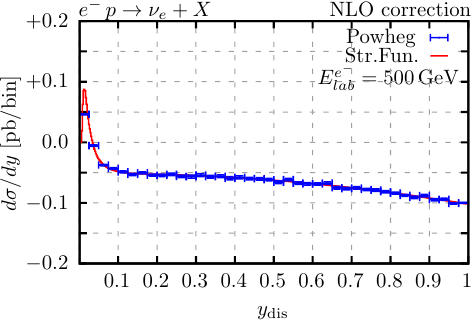}
  \includegraphics[width=0.45\textwidth,page=2]{pdfs/xbj.pdf}
  \hspace{1.25cm}
  \includegraphics[width=0.45\textwidth,page=2]{pdfs/ydis.pdf}
  \caption{Same as figure~\ref{fig:CCnxy} for charged current $l^-$ ($l^+$) DIS.
  }
\label{fig:lCC}
\end{figure}

\begin{figure}
  \includegraphics[width=0.45\textwidth,page=6]{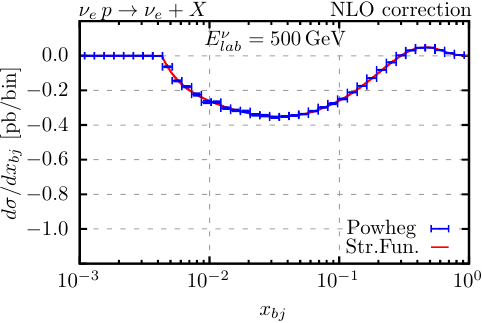}
  \includegraphics[width=0.45\textwidth,page=6]{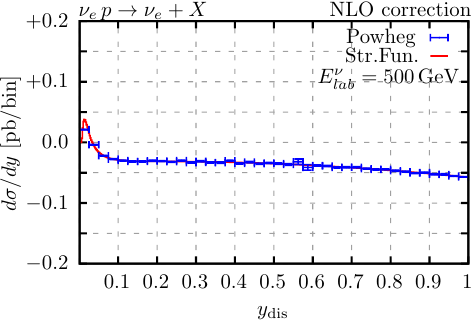}
  \includegraphics[width=0.45\textwidth,page=3]{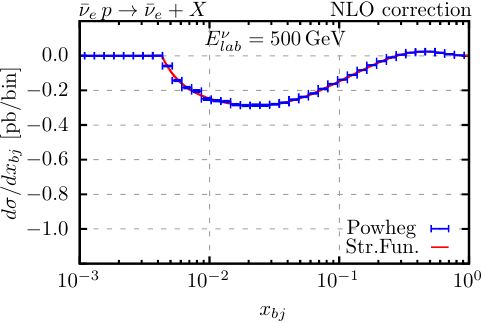}
  \hspace{1.25cm}
  \includegraphics[width=0.45\textwidth,page=3]{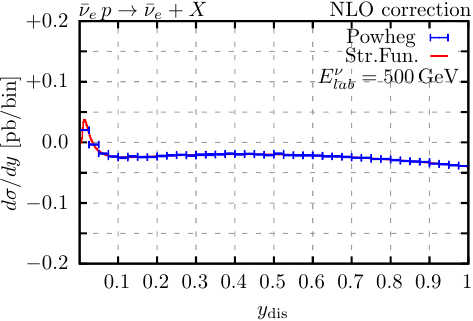}
  \caption{Same as figure~\ref{fig:CCnxy} for NC $\nu_e$ ($\bar{\nu}_e$) DIS.
  }
\label{fig:nNC}
\end{figure}

\begin{figure}
  \includegraphics[width=0.45\textwidth,page=4]{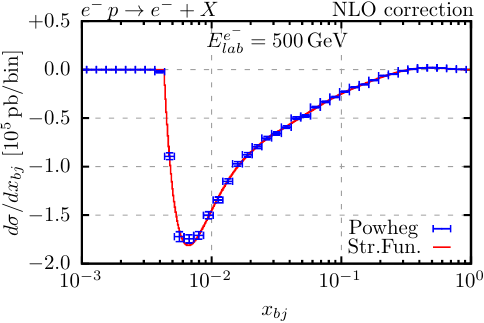}
  \includegraphics[width=0.45\textwidth,page=4]{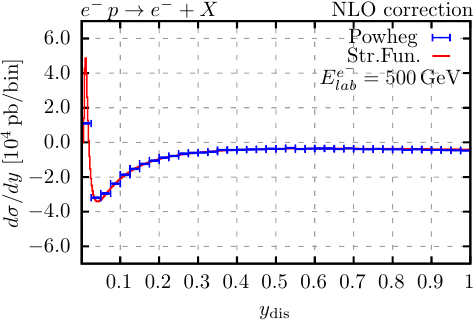}
  \includegraphics[width=0.45\textwidth,page=5]{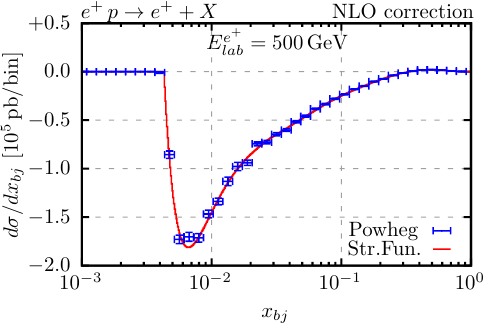}
  \hspace{1.25cm}
  \includegraphics[width=0.45\textwidth,page=5]{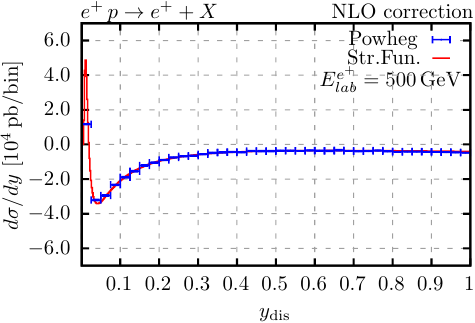}
  \caption{Same as figure~\ref{fig:CCnxy} for NC $l^-$ ($l^+$) DIS.
  }
\label{fig:lNC}
  \end{figure}

\bibliographystyle{JHEP}
\bibliography{pwhgvDIS}
\end{document}